\documentclass{iopart}

\pdfoutput=1

\usepackage{graphicx,epsfig}
\usepackage{bm}

\def\be{\begin{equation}}
\def\ee{\end{equation}}

\def\bea{\begin{eqnarray}}
\def\eea{\end{eqnarray}}

\def\a{\alpha}
\def\s{\sigma}

\begin{document}

\title[Entanglement  of two disjoint blocks in XY chains]
{Entanglement entropy of two disjoint blocks in XY chains}

\author{Maurizio Fagotti and Pasquale Calabrese}
\address{Dipartimento di Fisica dell'Universit\`a di Pisa and INFN, Pisa, Italy.}

\date{\today}

\begin{abstract}

We study the R\`enyi entanglement entropies of two disjoint intervals in XY chains.
We exploit the exact solution of the model in terms of free Majorana fermions and we show how to 
construct the reduced density matrix in the spin variables by taking properly into account the 
Jordan-Wigner string between the two blocks. 
From this we can evaluate any  R\`enyi entropy of finite integer order. 
We study in details critical XX and Ising chains and we show that the asymptotic results for large blocks 
agree with recent conformal field theory predictions if corrections to the scaling are included in the analysis
correctly.
We also report results in the gapped phase and after a quantum quench.

\end{abstract}

\maketitle

\section{Introduction}

The bipartite entanglement for a given division of the Hilbert
space into a part A and its complement B, can be measured in terms of the R\'enyi entropies \cite{Renyi} 
\be
S_A^{(\a)}=\frac1{1-\a}\log{\rm Tr}\,\rho_A^\a\,,
\ee
where $\rho_A={\rm Tr}_B\,\rho$ is the reduced density matrix of
the subsystem A, and $\rho=|\Psi\rangle\langle\Psi|$ is the
density matrix of the whole system in a pure state $|\Psi\rangle$.
The knowledge of the $S_A^{(\a)}$ for different $\a$ characterizes the
full spectrum of non-zero eigenvalues of $\rho_A$ 
\cite{cl-08}, and gives more information about the entanglement
than the more commonly studied von Neumann entropy $S_A^{(1)}$. It also
gives a fundamental insight into understanding the convergence and
scaling of algorithms based on matrix product states \cite{mps}.

In \cite{Holzhey,cc-04,cc-rev} it has been shown that for a one-dimensional
critical system whose scaling limit is described by a conformal
field theory (CFT), in the case where A is an interval of length $\ell$ 
in an infinite system, the asymptotic behavior of the quantities determining the R\'enyi
entropies is 
\begin{equation}\label{Renyi:asymp}
\Tr\rho_A^{\a}
\simeq c_\a \left(\frac{\ell}{a}\right)^{c(\a-1/\a)/6}\,,
\end{equation}
where $c$ is the central charge of the underlying CFT.
Thus the R\'enyi entropies (and in particular the von Neumann one for $\a=1$) give one of the best 
way of detecting the value of the central charge. 

Less attention has been devoted until now to the entanglement of two disjoint intervals in a 
CFT (and also in massive theories). It turned out that entanglement of disjoint intervals
is sensitive to universal details of the CFT that are not encoded in the central charge and it is
connected with the full spectrum of operators of the CFT underlying the lattice model.

\begin{figure}[b]
\includegraphics[width=0.6\textwidth]{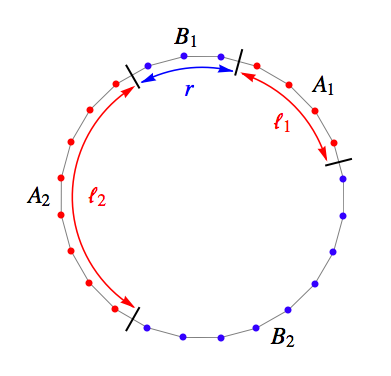}
\caption{Typical bipartition we consider in this paper. The subset $A$ is the union of two disjoint intervals 
$A_1$ and $A_2$ of length $\ell_1$ and $\ell_2$ respectively. The block separating them is denoted 
with $B_1$ of length $r$. The `environment' is $B=B_1\cup B_2$. The thermodynamic limit is obtained 
by sending the total length $L\to \infty$, while $\ell_1,\ell_2,r$ remain finite (i.e. the length of $B_2$ 
goes to $\infty$). }
\label{chain}
\end{figure}

We consider here the case of two disjoint intervals 
$A=A_1\cup A_2=[u_1,v_1]\cup [u_2,v_2]$ depicted in Fig. \ref{chain}. 
By global conformal invariance, in the thermodynamic limit,  $\Tr \rho_A^\a$ can be always written as
\be\fl
\Tr \rho_A^\a
=c_\a^2 \left(\frac{|u_1-u_2||v_1-v_2|}{|u_1-v_1||u_2-v_2||u_1-v_2||u_2-v_1|} 
\right)^{\frac{c}6(\a-1/\a)} F_{\a}(x)\,,
\label{Fn}
\ee 
where $x$ is the four-point ratio (for real $u_j$ and $v_j$, $x$ is real) 
\begin{equation}
x=\frac{(u_1-v_1)(u_2-v_2)}{(u_1-u_2)(v_1-v_2)}\,.
\label{4pR}
\end{equation}
Normalizing such that $F_\a(0)=1$, we have that $c_\a$ is the same non-universal constant 
appearing in Eq. (\ref{Renyi:asymp}).
The {\it universal} function $F_\a(x)$ depends explicitly on the full operator content 
of the theory and must be calculated case by case.
Originally it was proposed that $F_\a(x)=1$ identically \cite{cc-04}. 
This erroneous prediction has been tested in free fermion theories \cite{ch-08,ffip-08,kl-08}, 
and only the calculations in 
more complex theories allowed to detect this error \cite{cg-08,fps-09,cct-09,atc-09}.

Furukawa,  Pasquier, and  Shiraishi \cite{fps-09} calculated
$F_2(x)$  for a free boson compactified on a circle of radius $R$
\be
F_{2}(x)=
\frac{\theta_3 (\eta \tau) \theta_3 (\tau/\eta)}{ [\theta_3 (\tau)]^{2}},
\label{F2}
\ee
where $\tau$ is pure-imaginary, and is related to $x$ via
$x= [\theta_2(\tau)/\theta_3(\tau)]^4$. 
$\theta_\nu$ are Jacobi theta functions.
$\eta$ is a universal critical exponent dependent on the compactification radius $R$
(in Luttinger liquid literature $\eta=1/(2K)$).
This has been extended to general integer $\a\geq 2$ in Ref. \cite{cct-09}
\begin{equation}
F_\a(x)=
\frac{\Theta\big(0|\eta\Gamma\big)\,\Theta\big(0|\Gamma/\eta\big)}{
[\Theta\big(0|\Gamma\big)]^2}\,,
\label{Fnv}
\end{equation}
where $\Gamma$ is an $(\a-1)\times(\a-1)$ matrix obtained in \cite{cct-09},
$\eta$ is the same as above, while $\Theta$ is the Riemann-Siegel theta function
\begin{equation}
\label{theta Riemann def}
\Theta(0|\Gamma)\,\equiv\,
\sum_{m \,\in\,\mathbf{Z}^{\a-1}}
\exp\big[\,i\pi\,m^{\rm t}\cdot \Gamma \cdot m\big]\,.
\end{equation}
The XX model considered in the following is described by the compactified boson with $\eta=1/2$. 
We mention that the analytic continuation of this result to real $\a$ for general values of 
$\eta$ and $x$ (in order to obtain the entanglement entropy) is still an open problem, but the results for 
$x\ll1$ and $\eta\ll1$ are analytically known \cite{cct-09}. 

Nowadays, the only other example of non-trivial $F_\a(x)$ known exactly
is $F_2(x)$ for the Ising model that has the rather simple expression \cite{atc-09}
\be\fl
F_2(x)=\frac1{\sqrt{2}}\Bigg[
\left(\frac{(1 + \sqrt{x}) (1 + \sqrt{1 - x})}2\right)^{1/2} + x^{1/4} + ((1 - x) x)^{1/4} + (1 - x)^{1/4} \Bigg]^{1/2}.
          \label{CFTF2}
\ee
We mention that all these new results for entanglement entropy were derived by using some 
old `classical' results about CFT on orbifolds \cite{Dixon}.

It is worth to recall that in the case of many intervals, the 
entanglement entropy measures only the entanglement of the 
disjoint intervals with the rest of the system. 
It is {\it not} a measure of the entanglement of one interval with respect 
another, that instead requires the introduction of more complicated 
quantities because $A_1\cup A_2$ is in a mixed state (see e.g. Refs. 
\cite{Neg,Neg2,Neg3} for a discussion of this and examples). 

In this manuscript we report the exact evaluation of the R\`enyi entropies for integer $\a$ in XY spin 
chains with periodic boundary conditions, whose Hamiltonian is given by
\be
H_{XY}=-\sum_{l=1}^N\Bigl[\frac{1+\gamma}{4}\s_l^x\s_{l+1}^x+\frac{1-\gamma}{4}\s_l^y\s_{l+1}^y+\frac{h}{2}\s_l^z\Bigr]\, ,
\label{HXY}
\ee
where $\s_l^\a$ are the Pauli matrices at the site $l$, $h$ is the magnetic field and $\gamma$ is the 
so-called anisotropy parameter. 
For $\gamma=1$ the Hamiltonian reduces to the Ising model, while for $\gamma=0$
to the XX model. Hamiltonian (\ref{HXY}) can be diagonalized in terms of free-fermions (see below) but the 
fundamental difficulty is that the reduced density matrix of spin and fermion variables are not the 
same \cite{atc-09,ip-09}. While calculating the fermion matrix is straightforward, we derive here a 
general method to deal with spin reduced density matrices  and present actual calculations for 
the Hamiltonian (\ref{HXY}).

The paper is organized as follows. 
In the next section \ref{rdm} we give a general description of the density matrix mapping and 
we introduce our formalism. 
In section \ref{rensec} we derive the expressions for the R\'enyi entropies of integer order. 
These two sections are rather technical and are intended only for those readers that want to repeat
the calculations. The readers interested only in the results can skip directly to
section \ref{XXsec} where we report our results for the XX chain and compare them with CFT results by 
properly detecting corrections to the scaling.
In section \ref{Issec} we present the same analysis for the Ising model and we check the universality while 
changing $\gamma$.
In section \ref{NCsec} we consider non-critical chains.
In section \ref{quesec} we consider the R\'enyi entropies after a quantum quench.
Finally in section \ref{disc} we summarize our main results and we discuss topics deserving further 
investigations.
Four appendices contain technical details of the calculations and some background material.

\section{The reduced density matrix}
\label{rdm}

Generically a spin $1/2$ chain can be described by means of the Pauli matrices $\s_i^\mu$ 
with $\mu=0,1,2,3$ ($\s^0=1$ and $1\equiv x,\, 2\equiv y,\, 3\equiv z$) and $i$ labels the lattice sites. 
We are interested in the entanglement of a subsystem $A=\bigcup_{j=1}^N A_j$  consisting of $N$ 
disjoint spin blocks $A_j=[u_j,v_j]$. 
The reduced density matrix (RDM) can be written as a sum over all operators belonging to the blocks
forming $A$
\be\label{rdms}
\rho_A=\frac{1}{2^{|A|}}
\sum_{\mu_l}
\Bigl< \prod_{l\in A}\s_l^{\mu_l}\Bigr>\prod_{l\in A}\s_l^{\mu_l}\, ,
\ee
where $|A|=\sum_j \ell_j$ is the sum of the lengths of the blocks $\ell_j$.
The expectation value is taken over the state we need to calculate the entanglement. 
The knowledge of multi-point correlation functions $\Bigl< \prod_{l\in A}\s_l^{\mu_l}\Bigr>$ is however 
a very hard task even for eigenstates of integrable models (see e.g. \cite{afc-09,xxz} for the XXZ chains) 
and the above formula is particularly useful only when the spin-chain admits a representation in 
terms of free fermions, for which the Wick theorem suffices to calculate all of them.

The mapping between spin variables and fermion ones is obtained via the
Jordan-Wigner transformation (that for the XY chains makes the Hamiltonians quadratic) 
\be
c_l=\Bigl(\prod_{m<l}\sigma_m^z\Bigr)\frac{\sigma_l^x-i\sigma_l^y}{2}, \quad c_l^\dag=\Bigl(\prod_{m<l}\sigma_m^z\Bigr)\frac{\sigma_l^x+i\sigma_l^y}{2}\,.
\ee
It is also useful to define the Majorana fermions \cite{Vidal}
\be
a_{2l}=c^\dag_l+c_l\,,\qquad a_{2l-1}=i(c_l-c_l^\dag)\,,
\ee
satisfying anti-commutation relations $\{a_l,a_n\}=2\delta_{ln}$.
For the case of a single interval, a fundamental observation \cite{Vidal} is that the Jordan-Wigner string 
$\Bigl(\prod_{m<l}\sigma_m^z\Bigr)$ maps the space of the first $\ell$ spins in the space of the first
$\ell$ fermions and thus the RDMs of the first $\ell$ spins and fermions are the same, i.e.
\be
\rho_A=\frac{1}{2^\ell}
\sum_{\mu_l=0,1}
\Bigl< \prod_{l=1}^{2\ell} a_l^{\mu_l}\Bigr>\left(\prod_{l=1}^{2\ell} a_l^{\mu_l}\right)^\dag\, ,
\ee
with $a_l^{0}=1$ and $a_l^1=a_l$. 
At this point, it is worth to remind that the above equation gives directly all the eigenvalues of 
$\rho_A$ and so all R\`enyi entropies (see e.g. the reviews \cite{gl-rev} for more details). 
Indeed, one recognizes $\rho_A$ as an exponential form 
\be
\rho_A\propto e^{a_l W_{lm} a_m/4}\,,
\label{quad}
\ee
and using the Wick theorem (see also \ref{Gamma3})
\be
\tanh\frac{W}2=\Gamma, \qquad {\rm with}\;\; \Gamma_{ij}=\langle a_i a_j\rangle-\delta_{ij}\,.
\ee 
The $2\ell$ eigenvalues of  the correlation matrix $\Gamma$ have the form $\pm \nu_i$ with $\nu_i\in [-1,1]$, so one has all the $2^{\ell}$ eigenvalues of $\rho_A$ as 
\be
\lambda_{\{s\}}=\prod_i\frac{1+s_i \nu_i}{2}\, ,
\ee
for the $2^\ell$ possible choices of the variables $s_i=\pm1$. 
The R\`enyi entropies for any complex $\a$ are then given by
\be
S_A^{(\a)}=
\frac1{1-\a} \sum_{j=1}^{\ell} \ln \left[\left(\frac{1+\nu_j}2\right)^\a+ \left(\frac{1-\nu_j}2\right)^\a\right]\,.
\ee 
The calculation of $\nu_i$ only requires the diagonalization of a $2\ell\times 2\ell$
matrix, that is an extremely easy task compared to the diagonalization of the full 
$2^\ell\times 2^\ell$ density matrix. 
This expression can be further analytically massaged to obtain exactly the non-universal constant 
$c_\a$ \cite{jk-04} and also universal corrections to the scaling \cite{ccen-10}.

This is no longer true in the case of more disjoint spin-blocks. 
In the fermionic space, the subspace of two disjoint spin-blocks 
is not a local quantity: the non-locality of Jordan-Wigner transformation causes the 
fermions between the two blocks (i.e. those in $B_1$) to contribute to expectation values. 
For example, considering the product $\s_l^y\s_n^y$ with $l\in A_1$ and $n \in A_2$ and calling $r$ 
the distance between the blocks (see Fig. \ref{chain}), we have
\be\fl
\s_l^y\s_n^y=ia_{2l}\prod_{j=l+1}^{\ell_1}(ia_{2j-1}a_{2j})\Bigl[\prod_{j=\ell_1+1}^{\ell_1+r}(ia_{2j-1}a_{2j})\Bigr]\prod_{j=\ell_1+r+1}^{n-1}(ia_{2j-1} a_{2j})a_{2n-1}\,,
\ee
that depends on the `unpleasant' string of Majorana operators 
\begin{equation}\label{eq:S}
S\equiv \prod_{j=\ell_1+1}^{\ell_1+r}(ia_{2j-1}a_{2j})\, .
\end{equation} 
The string contribution cannot be neglected (as also shown in \cite{atc-09,ip-09}) and so 
the spin representation is not equivalent to the fermionic one.

For practical reasons we limit ourselves to the case of two blocks $|A_1|=\ell_1$ and $|A_2|=\ell_2$. 
We denote with $|B_1|=r$ the space between the two blocks, see Fig. \ref{chain} for a 
graphical representation. 
We focus on spin-chains with the ground-state property that only 
even numbers of fermionic operators have non-vanishing mean values, as e.g. those with an 
Hamiltonian commuting with $\prod_j\s_j^z$. 
In this case, in Eq. (\ref{rdms}) the product of operators $\prod \s^{\mu_l}_l$ does not depend on the 
string if we have an even number of Majorana operators (i.e. an even number of $\s^{x,y}$) 
in each block and it does depend if in each block we have 
an odd number of Majorana fermions (odd-even configurations give an odd total number of Majorana 
fermions that have vanishing expectation value by hypothesis).
Thus, the density matrix splits in two sums extended only over even and odd number of Majorana 
fermions in each block, i.e.
\bea
\label{eq:rhofer}
\rho_{A_1\cup A_2}&=&
\frac{1}{2^{\ell_1+\ell_2}}
\Bigl[ \sum_{\stackrel{\mu_l=0,1}{\rm even}}  \bigl<\prod_{l\in A_1} a_l^{\mu_l}\prod_{l\in A_2} a_l^{\mu_l}\bigr> 
                           \left(\prod_{l\in A_1} a_l^{\mu_l}\prod_{l\in A_2} a_l^{\mu_l}\right)^\dag\nonumber\\&&+
          \sum_{\stackrel{\mu_l=0,1}{\rm odd}}  \bigl<\prod_{l\in A_1} a_l^{\mu_l}S \prod_{l\in A_2} a_l^{\mu_l}\bigr> 
                           \left(\prod_{l\in A_1} a_l^{\mu_l}S\prod_{l\in A_2} a_l^{\mu_l}\right)^\dag \Bigr]
  \\     &\equiv& \frac{1}{2^{\ell_1+\ell_2}}\Bigl[ \sum_{\rm even}  \bigl< O_1  O_2\bigr> O_2^\dag  O_1^\dag
            +\sum_{\rm odd}  \bigl< O_1 S O_2\bigr> O_2^\dag S O_1^\dag\bigr]\,,
\eea
where $S$ is the complete string of $\s_z$ belonging to the region between the blocks (\ref{eq:S}).
We repeat that the two sums are intended over all possible products of Majorana fermions belonging 
to each interval and even/odd refers to the number of $\mu_l=1$, i.e. signaling the presence of a Majorana
operator. We introduced also the `short' $O_{1,2}$ for a general product of Majorana operators belonging 
to $A_{1,2}$. 
Eq. (\ref{eq:rhofer}) is an immediate consequence of the fermionic representation of spin variables and 
of the considerations above:
the string $S$ appears only in operators with odd number of $\s^{x(y)}$ in each block. 

Spin variables on different sites commute, and so $[ S,O_{1,2}]= 0$. Thus 
the spin RDM can be written as 
\be
\rho_{A_1\cup A_2}=\frac{1+ S}{2}\rho_+ +\frac{1- S}{2}\rho_-\, ,
\ee
where $\rho_{\pm}$ are the fermionic RDMs
\be\fl
\label{eq:rhopm}
\rho_{\pm}=\frac{1}{2^{\ell_1+\ell_2}}
\Bigl[ \sum_{\rm even} \bigl<O_1O_2\bigr>O_2^\dag O_1^\dag \pm 
\sum_{\rm odd} \bigl< O_1 S O_2\bigr> O_2^\dag  O_1^\dag\Bigr]\equiv \rho_{\rm even}\pm \rho_{\rm odd}\,.
\ee 
We will be interested in the R\`enyi entropies and in writing them in terms of the matrices $\rho_\pm$.
The orthogonal projectors $(1\pm  S)/2$ commute with $\rho_\pm$ and so 
the $n$-{th} power of the spin RDM is the following combination of the $n$-{th} powers of  fermionic RDMs
\be
\rho^n_{A_1\cup A_2}=\frac{1+S}{2}\rho_+^n+\frac{1-S}{2}\rho_-^n\, .
\ee
We note that $\rho_+$ is unitary equivalent to $\rho_-$, (the complete string 
$S_1$ (or $S_2$)  of $\s_z$ belonging to $A_1$ ($A_2$) is the unitary operator mapping one into 
the other, i.e. $S_1\rho_+S_1=\rho_-$).
As a consequence we have
\begin{equation}
\label{eq:spintoferm}
\Tr{\rho_{A_1\cup A_2}^n}=\Tr \rho_\pm^n\, ,
\end{equation}
where we used $\Tr S=0$. Still the matrices $\rho_\pm$ are {\it not} of the form (\ref{quad}) that are the
only objects we are able to deal with. 
Some further manipulations are still needed to bring them in a useful form.

However, before continuing on the main road, we notice that $\rho_\pm$ are 
the RDMs in fermionic variables of the state
\be
|\Psi_\pm\rangle=\Bigl(\frac{1\pm S}{2}+\frac{1\mp S}{2}S_2\Bigr)|\Psi_0\rangle\, ,
\ee
where $|\Psi_0\rangle$ is the ground state. These are the ground states of the Hamiltonians
\be
H_\pm=\Bigl(\frac{1\pm S}{2}+\frac{1\mp S}{2}S_2\Bigr)H\Bigl(\frac{1\pm S}{2}+\frac{1\mp S}{2}S_2\Bigr)\, .
\ee
We remark this unitary transformation is nontrivial only for operators on the contact surface between the blocks and the remaining chain.
Summarizing we can write a fermionic RDM equivalent to the spin RDM at the price of 
adding a finite number of  non local terms to the Hamiltonian $H$.

Now we write the fermionic RDM $\rho_\pm$ in Eq. (\ref{eq:rhopm}) as linear combinations 
of quadratic RDMs of the form (\ref{eq:rhopm}) that are the ones we are able to deal with. 
The basic intuitive reason why this is indeed possible is that
the $\s_z$ string is the exponential of a quadratic form
\be\fl
\prod_{l}\sigma_l^z=\prod_{l}(2c^\dag_lc_l-1)=\exp\Bigl(i\pi\sum_{l}c_l c^\dag_l\Bigr)=
\exp\Bigl( \frac{\pi}2 \sum_j (a_{2j-1}a_{2j}+i)\Bigr)\,,
\ee
and, in the case of a quadratic Hamiltonian, it can be interpreted as a term added to the exponential 
factor $W_{nm}$ in Eq. (\ref{quad}). 
These observations show that Wick theorem applies in some form for Eq. (\ref{eq:rhopm}), but the 
2-point correlations of $\rho_{\rm odd}$ are different from the 2-point correlations of $\rho_{\rm even}$.

The RDM's part $\rho_{\rm odd}$ includes multipoint correlations in which the string $S$ is inserted. 
To deal properly with this kind of objects we introduce the operator (that can be thought as a fake density 
matrix)
\be\label{rhoSA}
\rho^{S}_A=\frac{\Tr_B [S |\Psi \rangle\langle \Psi|]}{\langle S \rangle}\,,
\ee
where $|\Psi \rangle\langle \Psi|$ is the density matrix of the full system and the denominator is added 
to ensure the normalization $\Tr_A \rho_A^S=1$. This fake density matrix could be introduced for any 
operator, but in the following we need only to consider the string $S$. If $O$ is an operator acting on $A$,
$\rho_A^S$ satisfies the property $\Tr_A[\rho_A^S O]=\langle O S\rangle/ \langle S \rangle$. 
If the operator $S$ commutes with any operator with support in $A$, then 
$\rho_A^S$ is hermitian.

In Eq. (\ref{eq:rhopm}) the sums are done only on even and odd numbers of Majorana operators in each 
block. We need to write these sums in terms of all Majorana fermions. A useful observation
is that if we change sign to all the $a_l$ with $l\in A_1$ (or $l\in A_2$) in the multipoint correlators, then
$O_1$ ($O_2$) changes sign in the odd sum, but not in the even one. 
By taking the appropriate combination of 
them we are left with a sum over all possible Majorana fermions and not only over the even/odd ones. 
The operator that makes this useful change of sign is the Jordan-Wigner string $S_1$ ($S_2$) 
restricted to the first (second) interval, in fact using also $S_1^{-1}=S_1$, we have
\be
S_1 a_l S_1=
\cases{
-a_l\,,\qquad  l\in A_1\,,\cr
a_l\, ,\qquad   l\not\in A_1\,.
}
\ee
Thus we finally arrive to (${\bf 1}$ is the identity matrix)
\be
\rho_{\rm even}=\frac{\rho_A^{\bf 1}+S_1\rho_A^{\bf 1} S_1}{2}\,,
\qquad
\rho_{\rm odd}=\frac{\rho_A^S-S_1\rho_A^S S_1}{2}\,,
\ee
and so
\be
\rho_\pm=\frac{\rho_A^{\bf 1}+S_1\rho_A^{\bf 1} S_1}{2}\pm \frac{\rho_A^S-S_1\rho_A^S S_1}{2}\,.
\label{rhopm}
\ee
Notice that $\rho_A^{\bf 1}$ and $S_1\rho_A^{\bf 1} S_1$ have the same spectrum but 
different eigenvectors related by the matrix $S_1$. The same is true for the matrices  
$\rho_A^S$ and $S_1\rho_A^S S_1$, but they can have eigenvalues smaller than $0$ 
and so in no way can be seen as true density matrices.
In \ref{moreint} we generalized this form to the case of more disjoint intervals. 

Eq. (\ref{rhopm}) is the main result of this section: 
the rewriting of spin RDM as a linear combination of four fermionic RDM, i.e. exponential of a quadratic form
as in Eq.  (\ref{quad}).
These four matrices do {\it not commute} and  so they cannot be diagonalized simultaneously to 
find all eigenvalues of the spin RDM. 
However, if  we are interested in R\`enyi entropies with integer $\a$, we can handle 
this problem in a constructive way: we determine the product rules between RDMs and then we 
construct recursively any finite order R\`enyi entropy.  
Note that the RDM of the free-fermions is just given by $\rho_A^{\bf 1}$ in Eq. (\ref{rhopm}). 
This matrix can be simply diagonalized in the same way explained above for the single interval
as already done in Ref. \cite{ffip-08}.

\subsection{Product rule}

In this subsection we analyze the algebra of RDMs generated by a quadratic form, i.e. 
\begin{equation}
\label{eq:rhoW}
\rho_W=\frac{1}{Z(W)}\exp\Bigl(\Bigl.\sum_{l,n}a_l W_{l n} a_n\Bigr/4\Bigr)\,,
\end{equation}
where we do not assume $W$ to be hermitian (to include the non-hermitian contribution of the string). 
Anti-commutation relations of Majorana operators make always $W$ a complex skew-symmetric matrix, 
i.e. $W^T=-W$.
The constant 
\be
 Z(W)=\Tr {\exp\Bigl(\Bigl.\sum_{l,n}a_l W_{l n} a_n\Bigr/4\Bigr)}\,,
\ee 
ensures the normalization $\Tr \rho_W=1$.
In \ref{app:Skew-symmetry} we show that this normalization for complex diagonalizable 
skew-symmetric matrices is 
\be
Z(W)=\prod_{\{w\}/_\pm}2\cosh\Bigl(\frac{w}{2}\Bigr)\, ,
\ee
where $\{w\}/_\pm$ is the set of eigenvalues of $W$ with halved degeneration ($W$ is a skew-symmetric
matrix so any even function of $W$ has eigenvalues with even degeneracy). 
We assumed $Z(W)\neq0$, pathological cases can be cured as explained in \ref{app:Miscellanea}.

The product of fermionic RDMs of the form (\ref{eq:rhoW}) is
\be\label{PR}
\rho_W\rho_{W'}=\frac{Z\bigl(\log(\exp(W)\exp(W'))\bigr)}{Z(W)Z(W')}\rho_{\log(\exp(W)\exp(W'))}\, .
\ee
Indeed the commutator of operators in the exponent of (\ref{eq:rhoW}) 
\be
\sum_{l,n,j,k}\frac{W_{l n}W_{j k}'}{16} \Bigl[a_l a_n,a_j a_k\Bigr]=\frac{\vec a^T[W,W']\vec a}{4}\, .
\ee
is the essential ingredient in the Baker-Campbell-Hausdorff formula for the product of exponential of 
operators.

Fermionic RDMs are specified by correlation matrices
\be
\Gamma_{i j}=\Tr [a_i\rho_W a_j]-\delta_{i j}\,,
\ee
which can be written as (see \ref{app:Skew-symmetry}) 
\be\label{gammavsW}
\Gamma=\tanh\Bigl(\frac{W}{2}\Bigr)\quad \Longrightarrow \quad e^W=\frac{1+\Gamma}{1-\Gamma}\, .
\ee
Clearly the second equation is true only when $1-\Gamma$ is an invertible matrix. 
At this point, let us briefly summarize the logic of the following derivation. 
We can easily calculate/manipulate the correlation matrix $\Gamma$ that via Eq. (\ref{gammavsW}) 
gives the exponential factor $W$ that defines the quadratic density matrix in Eq. (\ref{eq:rhoW}).
We need to find what are the consequences of  the product rule of RDMs for the correlation matrices, i.e.
we need to find what is the correlation matrix corresponding to the product of two RDMs. 
While, through the chains of equations above, any $W$ defines a single $\rho_A$, the opposite is not 
true and there are several possible $W$'s for each $\rho_A$. 
Nevertheless, we can give a unique recipe for the composition of correlation matrices.

We indicate this matrix operation as $\Gamma\times \Gamma'$ (notice it is not the product of the matrices) 
and it is formally defined by  Eq. (\ref{PR}) as 
\be\label{comp}
\rho[\Gamma]\rho[\Gamma']=\Tr\left[\rho[\Gamma]\rho[\Gamma']\right]
\rho[\Gamma\times\Gamma']\, .
\ee
To specify this operation we still need two ingredients:
\begin{enumerate}
\item an usable expression for the correlation matrix 
\be
(\Gamma\times\Gamma')_{i j}=
\frac{Z(W)Z(W')}{Z\bigl(\log(\exp(W)\exp(W'))\bigr)}\Tr{a_i\rho_W\rho_{W'}a_j}-\delta_{i j}\,,
\ee
associated to the product $\rho_W\rho_{W'}\equiv\rho[\Gamma]\rho[\Gamma']$;
\item an expression for the trace of two fermionic RDMs
\be
\{\Gamma,\Gamma'\}\equiv \Tr \rho[\Gamma]\rho[\Gamma']
=\frac{Z(W)Z(W')}{Z\bigl(\log(\exp(W)\exp(W'))\bigr)}\,,
\ee
in terms of the correlation matrices $\Gamma$ and $\Gamma'$.
\end{enumerate}

The first requirement is easily obtained if we assume $1-\Gamma$ and $1-\Gamma'$ invertible. 
Indeed, if we make explicit the exponential products
\be
\frac{1+\Gamma\times\Gamma'}{1-\Gamma\times\Gamma'}=\frac{1+\Gamma}{1-\Gamma}\frac{1+\Gamma'}{1-\Gamma'}\, ,
\ee
after simple algebra we obtain
\begin{equation}\label{eq:GtimesGprime}
\Gamma\times\Gamma'=1-(1-\Gamma')\frac{1}{1+\Gamma\Gamma'}(1-\Gamma)\, .
\end{equation}
$\Gamma\times\Gamma'$ is a skew-symmetric matrix, even if it is not obvious from the above formula.
We checked that this relation remains true also if  $1-\Gamma$ is not invertible (at least for the 
kind of matrices we are interested in), but a complete rigorous proof of Eq. (\ref{eq:GtimesGprime}) 
is beyond the goal of this paper.

The second request is less trivial because the correlation matrix $\Gamma$ does not determine univocally 
the matrix $W$, and the sign of $Z(W)$ remains ambiguous. 
However,  $\{\Gamma,\Gamma'\}$ is a functional of $\Gamma$ and $\Gamma'$, i.e. it is  the product 
of the eigenvalues of $(1+\Gamma\Gamma')/2$ with halved degeneration 
(the spectrum of $\Gamma\Gamma'$ is double degenerate \cite{if-09})
\be
\{\Gamma,\Gamma'\}=\prod_{\mu\in \mathrm{Spectrum}[\Gamma\Gamma']/2}\frac{1+\mu}{2}=
\pm\sqrt{\det\Bigl|\frac{1+\Gamma\Gamma'}{2}\Bigr| }\, .
\ee
The unspecified $\pm$ sign in front is the ambiguity
that (as shown in \ref{app:Skew-symmetry}) 
can be solved by rewriting the composition rule as
\be
\{\Gamma,\Gamma'\}=\exp{\Bigl(\frac{1}{2}
\int_{\gamma_{[0\to 1]}}\frac{\mathrm{d}\lambda}{1+\lambda} \Tr{\frac{\Gamma\Gamma'-1}{\lambda\Gamma\Gamma'+1}}\Bigr)}\,.
\ee
that does not depend on the  curve $\gamma$.

It is evident that the operation $\times$ is associative and so we are in position to make any 
product of fermionic RDMs:
\begin{equation}\label{eq:prodrule}
\prod_{i=1}^n\rho[\Gamma_i]=\{\Gamma_1,\cdots,\Gamma_n\}\rho[\Gamma_1\times\cdots\times\Gamma_n]\, ,
\end{equation}
where
\be\fl\label{boh}
\{\Gamma_1,\Gamma_2,\Gamma_3\dots,\Gamma_n\}\equiv 
\Tr[\rho[\Gamma_1]\rho[\Gamma_2]\cdots]=
\{\Gamma_1,\Gamma_2\}\{\Gamma_1\times \Gamma_2,\Gamma_3\dots,\Gamma_n\}\, .
\ee
If, for some $i$ and $j$, the matrix $(1+\Gamma_i\Gamma_j)/2$ is not invertible, Eq. (\ref{eq:prodrule})
cannot be applied. It is still possible to isolate the pathological parts and we report the details of 
this procedure  in  \ref{app:Miscellanea}.

\section{R\'enyi  entropies} 
\label{rensec}

From equations (\ref{eq:spintoferm}), (\ref{rhopm}) and (\ref{eq:prodrule}), 
the R\`enyi entropies for integer $\a$ can be written as 
\begin{equation}
\label{eq:Salpha}
S_\a=\frac{1}{1-\a}\log\Bigl[
\frac{1}{2^\a}\sum_{\zeta_1,\dots,\zeta_\alpha}\prod_{i=1}^\a c[\zeta_i]\{\Gamma_{\zeta_1},
\cdots,\Gamma_{\zeta_{\a}}\}\Bigr]\, ,
\end{equation}
where we defined the variables $\zeta_i=1,2,3,4$ (that label which of the terms in Eq. (\ref{eq:Salpha}) 
is taken in the particular product) and we defined the shorts for the 2-point correlation matrices
\be\fl
\Gamma_1=\Gamma_{\rho^{\bf 1}}\,, \qquad
\Gamma_2=\Gamma_{S_1 \rho^{\bf 1} S_1}\,,\qquad
\Gamma_3=\Gamma_{\rho^S}\,, \qquad
\Gamma_4=\Gamma_{S_1 \rho^S S_1}\,,
\ee
and
\be
c[\zeta]=
\cases{
1& $\zeta\in\{1,2\}$,\cr
\langle{S}\rangle&$\zeta=3$\,,\cr
-\langle{S}\rangle&$\zeta=4$\,.
}
\ee

In the case of $\a=2$ there are no problems with singular terms (see \ref{app:Skew-symmetry}) and 
the above expression can be re-written as 
\be\fl
S_2=-\log\left[\frac{1}{4}\sum_{\zeta_1,\zeta_{2}}c[\zeta_1]c[\zeta_2](\pm)\sqrt{\det\Bigl|\frac{1-\Gamma_{\zeta_1}}{2}\frac{1-\Gamma_{\zeta_2}}{2}+\frac{1+\Gamma_{\zeta_1}}{2}\frac{1+\Gamma_{\zeta_2}}{2}\Bigr|}\right]\, ,
\ee
where here and in the following equation we leave the sign ambiguity unspecified. 
Taking into account the trace's invariance under cyclic permutations, $S_2$ becomes the logarithm of 
a sum of 10 terms
\bea\fl
e^{-S_2}&=&\frac{1}{4}\sum_{\zeta}c[\zeta]^2\sqrt{\det\Bigl|\frac{1+\Gamma_{\zeta}^2}2\Bigr|} +
\nonumber\\ \fl &&
+\frac{1}{2}\sum_{\zeta_1>\zeta_2}\!\!\!c[\zeta_1]c[\zeta_2](\pm)\sqrt{\det\Bigl|\frac{1-\Gamma_{\zeta_1}}{2}\frac{1-\Gamma_{\zeta_2}}{2}+\frac{1+\Gamma_{\zeta_1}}{2}\frac{1+\Gamma_{\zeta_2}}{2}\Bigr|}\, .
\eea 

These formulas are already usable for a direct computation of R\`enyi entropies. There are however 
some simplifications that occur by using the property of the correlation matrices $\Gamma$'s.
The first matrix $\Gamma_1$ is the standard fermionic correlation matrix (i.e. the one corresponding 
to free fermions in the absence of the Jordan-Wigner string, already considered in Ref. \cite{ffip-08}), 
$\Gamma_2$ can be obtained 
from $\Gamma_1$  as $\Gamma_2=P_1\Gamma_1 P_1$, with $P_1$ the Hermitian unitary matrix
\be
P_1\equiv
\left(\begin{array}{cc}
-{\bf 1}_{A_1}&0\\
0&{\bf 1}_{A_2}
\end{array}\right)\, ,
\ee 
The same relation occurs between the third and the forth matrix $\Gamma_4=P_1\Gamma_3 P_1$.
Instead  $\Gamma_3$ is not trivially related to $\Gamma_1$. 
In \ref{Gamma3} we prove the following identity 
\be\label{sh}
\Gamma_3=\Gamma_1-\Gamma_{A B_1}\Gamma_{B_1 B_1}^{-1}\Gamma_{B_1 A}\,,
\ee
where the double subscripts take into account restrictions to rectangular correlation matrices, 
i.e. the first (second) subscript identifies the region where the row (column) index runs. 
We show in  \ref{moreint} that similar properties are valid also for an arbitrary number of intervals.

Using these relations after some algebraic manipulations one can write down the full sums for the 
R\`enyi entropies. It is important to notice that $\langle S \rangle= {\rm Pf}(\Gamma_{B_1})$ and so 
$\langle S \rangle^2= \det(\Gamma_{B_1})$. Furthermore to short the notations we denote
\be
\{\cdots,\Gamma_i^n,\cdots\}=\{\cdots, \underbrace{\Gamma_i,\cdots,\Gamma_i}_n,\cdots\}\,.
\ee
Finally $S_2$ can be written in a rather simple way
\begin{equation}\label{eq:Renyidouble2}
S_2=-\log\Bigl[\frac{\{\Gamma_1^2\}+\{\Gamma_1,\Gamma_2\}}{2}+(\det{\Gamma_{B_1}})\frac{\{\Gamma_3^2\}-\{\Gamma_3,\Gamma_4\}}{2}\Bigr]\, .
\end{equation}
But, increasing the order $\a$, the explicit expressions become soon long. 
For example, we give the (simplified) formulae for $S_3$
\begin{equation}\label{eq:Renyidouble3}\fl
S_3=-\frac{1}{2}\log\Bigl[\frac{\{\Gamma_1^3\}+3\{\Gamma_1^2,\Gamma_2\}}{4}+3(\det\Gamma_{B_1})
\frac{\{\Gamma_1,\Gamma_3^2\}+\{\Gamma_2,\Gamma_3^2\}-2\{\Gamma_1,\Gamma_4,\Gamma_3\}}{4}\Bigr]
\end{equation}
 and $S_4$
\bea\fl
\label{eq:Renyidouble4}
S_4&=&-\frac{1}{3}\log\Bigl[\frac{\{\Gamma_1^4\}+4\{\Gamma_1^3,\Gamma_2\}+2\{\Gamma_1^2,\Gamma_2^2\}
+\{\Gamma_1,\Gamma_2,\Gamma_1,\Gamma_2\}}{8}+
\\ \fl&&\nonumber
+(\det\Gamma_{B_1}) \left(\frac{\{\Gamma_1,\Gamma_3,\Gamma_1,\Gamma_3\}+\{\Gamma_1,\Gamma_4,\Gamma_1,\Gamma_4\}}{4}+
\frac{\{\Gamma_1^2,\Gamma_3^2\}+\{\Gamma_1^2,\Gamma_4^2\}}{2}
\right.\\ \fl&&\nonumber\left.
\Bigl.+\frac{\{\Gamma_1,\Gamma_3,\Gamma_2,\Gamma_3\}-
\{\Gamma_3,\Gamma_1,\Gamma_4,\Gamma_1\}-
\{\Gamma_1,\Gamma_2,\Gamma_3,\Gamma_4\}
}{2}+\right.\\ \fl&&\nonumber\left. +\frac{
-\{\Gamma_1,\Gamma_3,\Gamma_2,\Gamma_4\}
-\{\Gamma_1,\Gamma_2,\Gamma_4,\Gamma_3\}}{2}+\{\Gamma_1,\Gamma_2,\Gamma_3^2\}
-\{\Gamma_3,\Gamma_4,\Gamma_1^2\}\right)
\\ \fl&&\nonumber
+
(\det \Gamma_{B_1})^2\frac{\{\Gamma_3^4\}-4\{\Gamma_3^3,\Gamma_4\}
+2\{\Gamma_3^2,\Gamma_4^2\}+\{\Gamma_3,\Gamma_4,\Gamma_3,\Gamma_4\}}{8}\Bigr]\, .
\eea

\section{Critical XX model}
\label{XXsec}

In this section we report the explicit results for the XX chain in zero magnetic field 
(i.e. $\gamma=0$ and $h=0$ in Eq. (\ref{HXY})).
XX chains have been previously analyzed by Furukawa et al \cite {fps-09} by means of exact 
diagonalization techniques that allowed to explore relatively small chains with at most $30$ spins. 
The asymptotic results from CFT (cf. Eq. (\ref{F2})) were obscured in this previous analysis by  
large oscillating corrections to the scaling. 
The smallness of the systems and the lack of a precise knowledge of the form of the correction to the 
scaling made impossible any finite-size scaling analysis to check the CFT predictions (\ref{F2}). 
In fact these oscillating corrections to the scaling have been widely observed in last years 
\cite{lsca-06,r-2,ar-10}, but a precise theory about their origin and their exact form is only 
recently available \cite{ccen-10,cc-10}.

\begin{figure}
\includegraphics[width=0.75\textwidth]{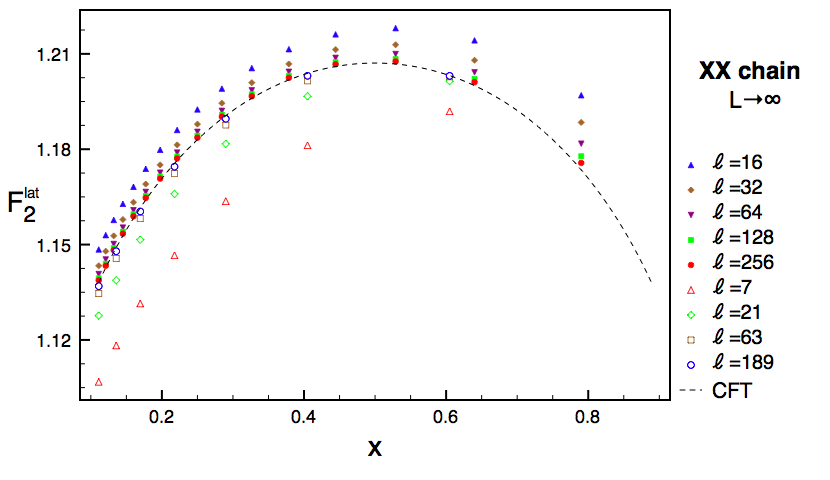}
\includegraphics[width=0.75\textwidth]{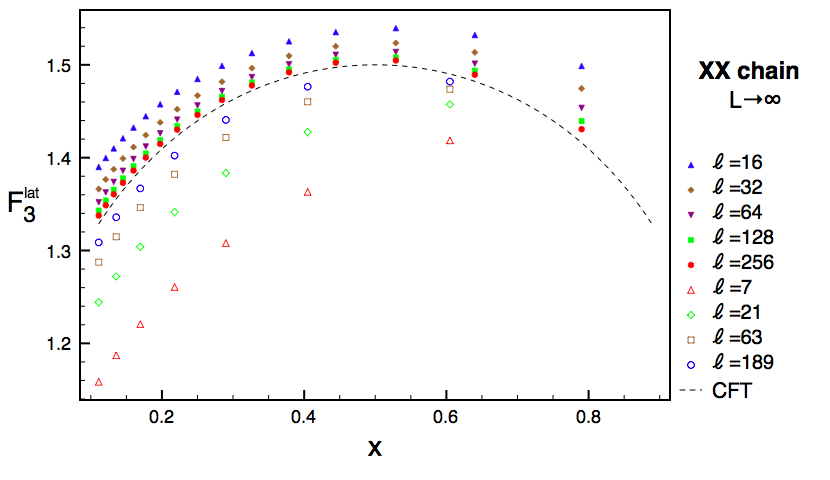}
\includegraphics[width=0.75\textwidth]{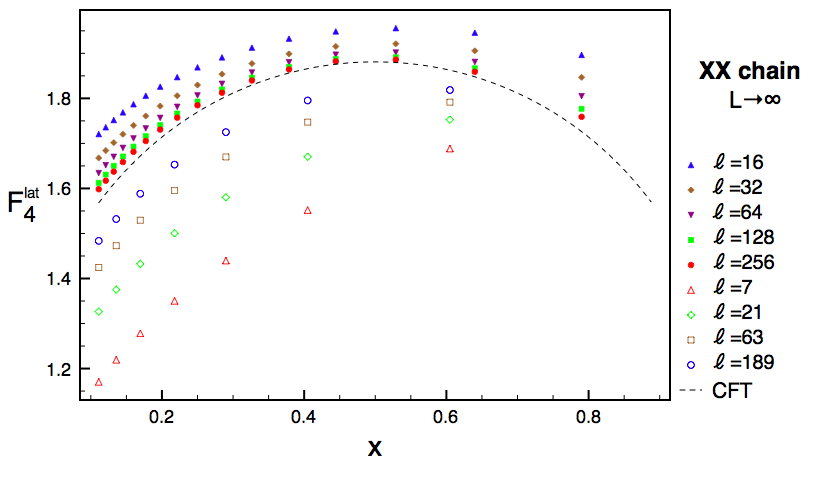}
\caption{Scaling function $F_\a^{\rm lat}(x)$ for $\a=2,3,4$ (from top to bottom) 
for the XX model in the thermodynamic limit and for various $\ell=\ell_1=\ell_2$ and $r$. 
Corrections to the scaling show even-odd oscillations with $\ell$ as in the single interval case. 
The results converge quickly to the universal CFT prediction $F_\a(x)$.
} 
\label{XXvsCFT}
\end{figure}

By exploiting the exact solution, we can avoid these problems and explore large 
enough values of $\ell$ allowing a finite-$\ell$ scaling analysis similar to the one for a single block 
\cite{ccen-10}. 
We start from the infinite volume limit. We consider two intervals both of length 
$\ell=\ell_1=\ell_2$ at distance $r$.
In this case, the four-point ratio $x$ in Eq. (\ref{4pR}) is given by
\be
x=\left(\frac{\ell}{\ell+r}\right)^2\,,
\label{4pRlat}
\ee
and the scaling of the quantity determining the R\'enyi entropies is (we use $c=1$)
\be
\Tr \rho_A^\a= 
c_\a^2\left(\frac{(\ell+r)^2}{\ell^2 r(2\ell+r)}\right)^{(n-1/n)/6}F_\a^{\rm lat}(x,\ell)\, ,
\ee
that is an implicit definition of $F_\a^{\rm lat}(x,\ell)$ encoding the fact that for finite $\ell$ we expect
corrections to the scaling of the form
\be
F_\a^{\rm lat}(x,\ell)=F_\a^{\rm CFT}(x)+\ell^{-\delta_\a}f_\a(x)+\dots\, ,
\ee
where $F_\a^{\rm CFT}(x)$ is the universal quantity appearing in Eq. (\ref{Fnv}).
The exponent $\delta_\a$ governs the leading correction to the scaling. It has been shown with CFT in 
Ref. \cite{cc-10} that this exponent is equal to $2x/\a$ independently of the number of intervals and $x$ is 
the scaling dimension of an operator introduced by the conical singularity necessary to describe the 
reduced density matrix. For a single interval, it has been shown by exact analytic calculation
that $\delta_\a=2/\a$ \cite{ccen-10} and thus we expect the same exponent for the double 
interval case (the presence of a relevant operator with dimension $x=1$ has been 
justified in Ref. \cite{cc-10}).
In the following analysis we will greatly benefit of the exact knowledge of the quantity $c_\a$
from the exact solution of the single interval entanglement \cite{jk-04,p-04} and so our results will be also 
a non-trivial check of the equivalence of the two non-universal multiplicative constants in the single and 
double interval case.

We report in Fig. \ref{XXvsCFT}  results for the function $F_{2,3,4}^{\rm lat}(x,\ell)$ 
for various $\ell$ and $x$. It is evident that irrespective of the value of $x$, with increasing 
$\ell$ the results approach the CFT prediction. For odd $\ell$ the
asymptotic result is approached from below, while for even ones it is approached from above. 
These are the already mentioned oscillations that made difficult the analysis based on small chains.
These plots are the first direct quantitative tests of the correctness of Eq. (\ref{Fnv}). 
These figures do not leave doubts about the correctness of the results of Ref. \cite{cct-09}.
It is worth to mention that, as already shown elsewhere \cite{fps-09,cct-09,atc-09}, the finite $\ell$
curves do not have the symmetry $x\to 1-x$ that is restored only in the $\ell\to \infty$ limit Eq. (\ref{Fnv}).

\begin{figure}
\includegraphics[width=0.55\textwidth]{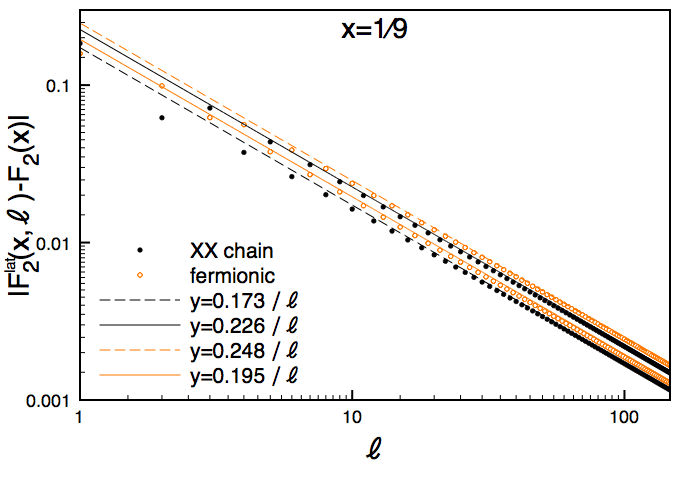}
\includegraphics[width=0.55\textwidth]{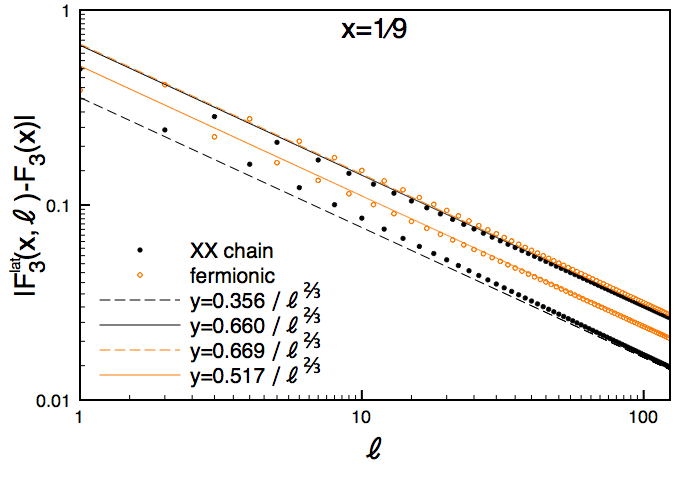}
\includegraphics[width=0.55\textwidth]{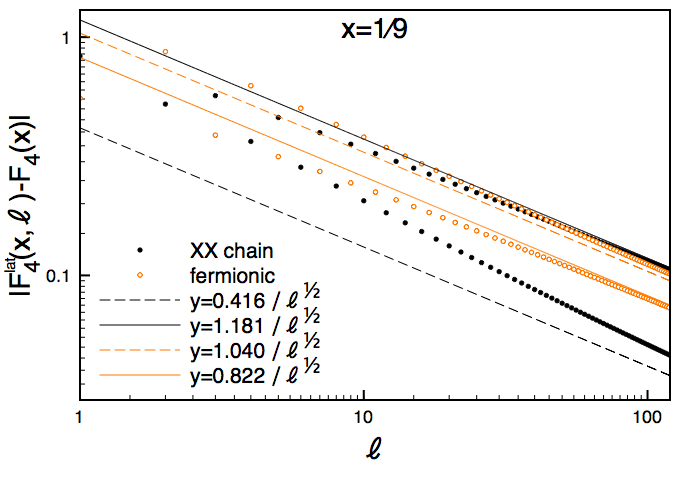}
\includegraphics[width=0.55\textwidth]{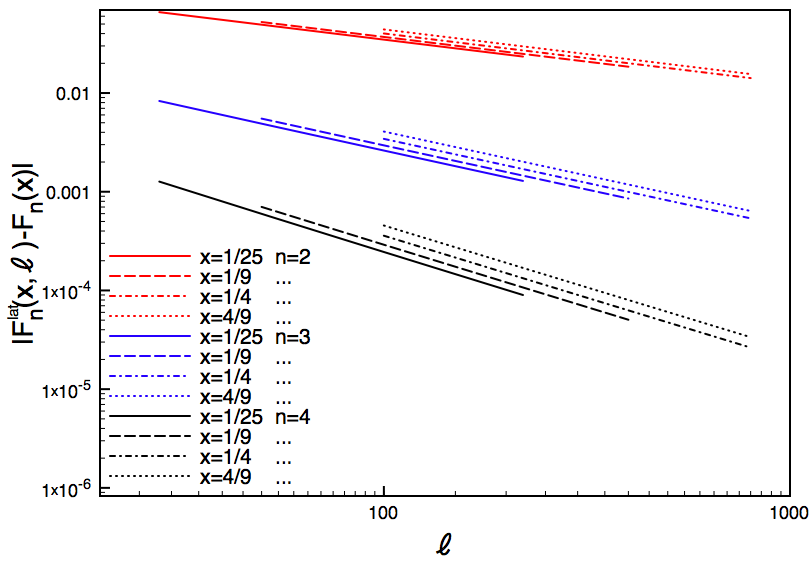}
\caption{Corrections to the scaling for $F_\a^{\rm lat}(x)$ obtained by subtracting the asymptotic value.
The first three panels show that for $\a=2,3,4$  at fixed $x=1/9$, both for fermionic and spin variables 
corrections to the scaling have exponents $\delta_\a=2/\a$, as for one interval. 
For fermionic degrees of freedom the asymptotic behavior is obtained for smaller values of $\ell$. 
Last panel: $x$ independence of the corrections to the scaling exponents. All curves with different $x$
at fixed $\a$ lie on parallel lines.} 
\label{corrXX}
\end{figure}

Having established the correctness of the asymptotic form, we can now move to the finite
$\ell$ corrections and check whether the prediction $\delta_\a=2/\a$ \cite{cc-10} is correct. 
Having precise control of the corrections to the scaling is not an academic task: their analysis is 
fundamental to provide accurate results when such large system sizes are not available (in previous 
studies they would have been an important tool) and in cases when the asymptotic form is not 
known (as in the following section for the Ising model for $\a>2$). 
In Fig. \ref{corrXX} we report the function $|F_\a(x)-F_\a^{\rm lat}(x)|$ for fixed $x=1/9$ and 
$\a=2,3,4$.
We both report results for the spin R\`enyi entropies and for the fermionic ones (i.e. without considering
the string contribution). We recall that for free fermions we have  $F_\a(x)=1$ identically.
The results show a power law behavior for large enough $\ell$ with the predicted exponent
$2/\a$ as for the single interval. Notice that by increasing $\a$, the values $\ell$ where the leading 
asymptotic correction can be identified become larger and larger, in analogy with the single block case
\cite{ccen-10} (as obvious because of the smallness of the exponent $\delta_\a$).
For the fermionic variables the asymptotic behavior is reached before than for spin
degrees of freedom: the string introduces further corrections to the scaling that in the present model
are subleading.
To show the $x$ independence of this exponent in the last panel of Fig. \ref{corrXX}, 
we report the same kind of plots for different values of $x$, showing that, at fixed $\a$, 
the corrections lie on parallel lines.

\begin{figure}
\includegraphics[width=\textwidth]{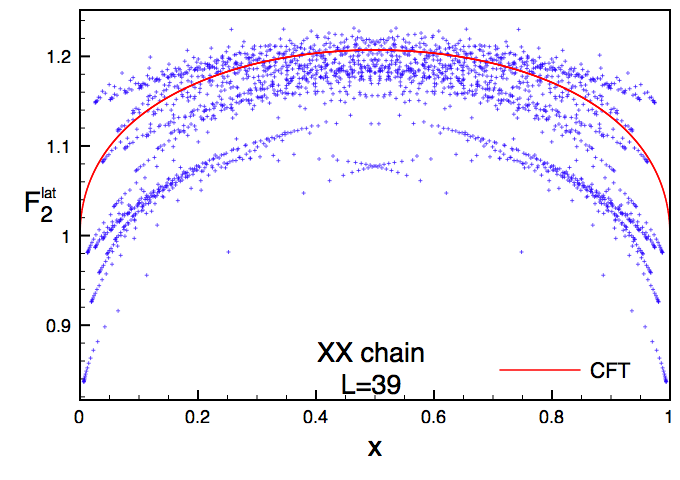}
\caption{The function $F_2^{\rm lat}(x)$ for a small chain with $L=39$ spins. 
Oscillating corrections to the scaling prevent to see the universal CFT prediction shown as a continuous 
curve.} 
\label{caosXX}
\end{figure}

To conclude this section we present some results for finite systems. 
It is known that all the formulas above (including the $x$ dependence) at the leading order in finite 
systems can be described by replacing any distance $u_{ij}$ by the chord length 
$u_{ij}\to \frac{L}{\pi}\sin\frac{\pi u_{ij}}L$. We checked that this rescaling is indeed correct and that, 
for large enough $\ell$ and $L$, the results agree with CFT. 
These plots give no further information compared to the ones already presented and we do not report them. 
We only show the results for a rather small system of length $L=39$ (that nevertheless is above 
anything obtainable by exact diagonalization).
In Fig \ref{caosXX} we report the resulting $F_2^{\rm lat}$ for {\it all} the possible divisions in four 
parts of this chain of length $L=39$.
For such small chain, the results are obviously very unclear since the corrections to the scaling 
are obscuring the CFT scaling represented by a continuous line that is surrounded by the points, 
signaling the oscillatory nature of the corrections (for clarity, compare with the analogous plot for the 
Ising model in next section).

\section{Critical Ising universality class}
\label{Issec}

\begin{figure}
\includegraphics[width=0.75\textwidth]{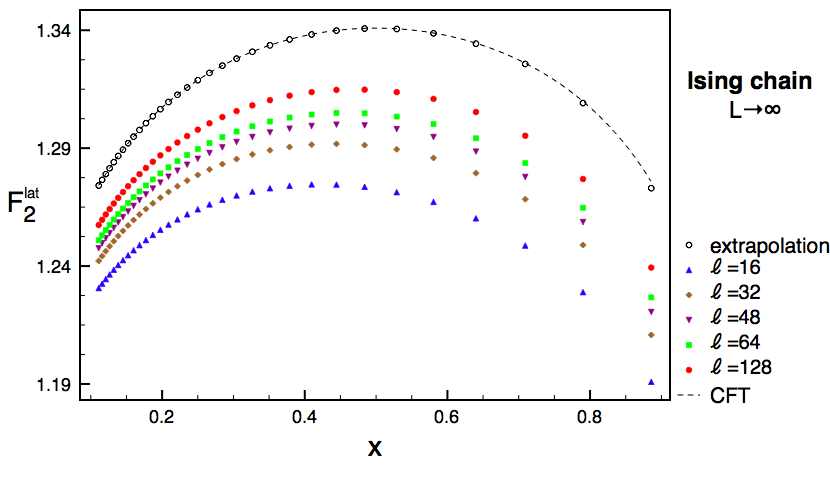}
\includegraphics[width=0.75\textwidth]{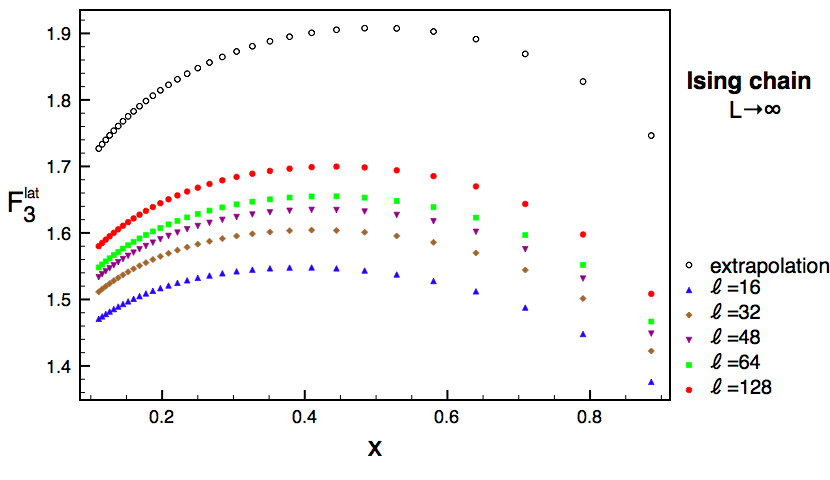}
\includegraphics[width=0.75\textwidth]{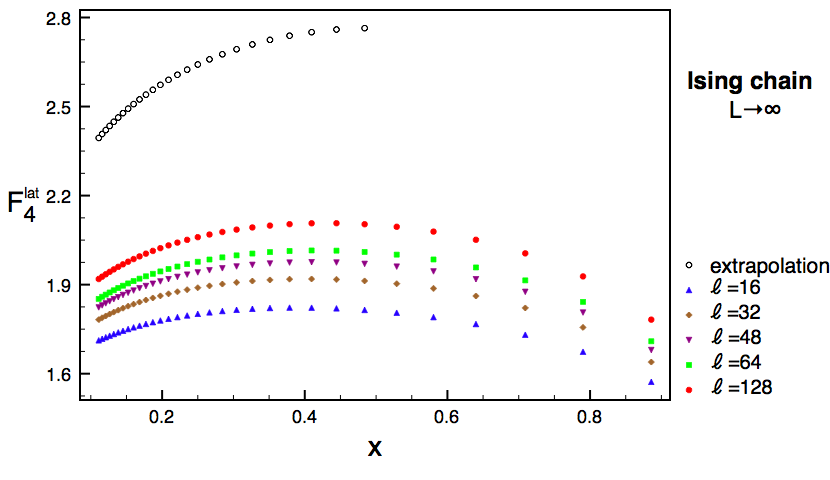}
\caption{Scaling function $F_\a^{\rm lat}(x)$ for $\a=2,3,4$ (from top to bottom) 
for the Ising model in the thermodynamic limit and for various $\ell$. 
Corrections to the scaling are monotonous. 
The top curve in each plot is the extrapolation to $\ell\to\infty$. 
The convergence to the universal CFT prediction $F_\a(x)$ is slower than in the
XX case, because the leading exponent of corrections to the scaling is $1/\a$.
} 
\label{ISvsCFT}
\end{figure}

The Ising model is given by Hamiltonian (\ref{HXY}) with $\gamma=1$ and it is critical for $h=1$. 
Results for this model have been already derived numerically for $\a=2$ in Ref. \cite{atc-09} by using
a tree tensor network algorithm and Monte Carlo simulations of the two-dimensional 
classical problem in the same universality class. 
These calculations allowed a precise determination of $F_2(x)$, but the system
sizes explored were not enough to analyze R\`enyi entropies with larger values of $\a$.
In the course of our analysis we always compared our data for small systems 
with those in Ref. \cite{atc-09}, in order to check the correctness of both methods.

\begin{figure}
\includegraphics[width=\textwidth]{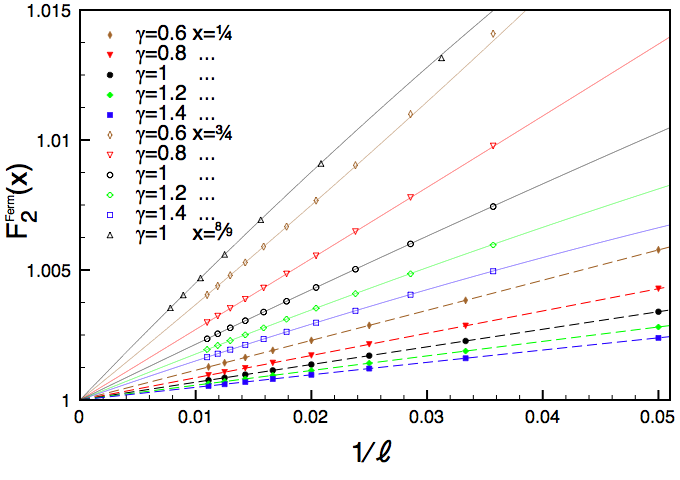}
\caption{$F_2^{\rm Ferm}(x)$ for  different $x$ and $\gamma$. 
All results present the same leading correction to the scaling exponent $\delta_2=1$. 
The extrapolated data at $\ell\to\infty$ collapse in the single point $F_2(x)=1$. 
} 
\label{corrIfer}
\end{figure}

We report our results in the thermodynamic limit for various values of $\ell=\ell_1=\ell_2$ at different 
separations $r$ (resulting in the four-point ratio $x$ given in Eq. (\ref{4pRlat})) in Fig. \ref{ISvsCFT}.
As already noticed in Ref. \cite{atc-09}, oppositely to the XX chain, in the Ising model we have monotonic 
finite $\ell$ correction to the scaling. For finite $\ell$, the results do not show the symmetry $x\to1-x$ valid 
for infinite $\ell$. This is restored only by the extrapolated data at $\ell\to\infty$. 
To perform this extrapolation in the most accurate way, we have first to determine the correction to the 
scaling exponent. For a single interval, it is exactly known that the leading corrections are characterized by
the exponent $\delta_\a=2/\a$ as for the XX chain (the origin of the equality \cite{ccen-10} is due 
to the equivalence of the two RDMs \cite{ij-08}). 
However for $\a=2$, it has been already shown \cite{atc-09} that $\delta_2=1/2$,
different from the single interval one. 
This result is quite surprising also in view the recent CFT 
analysis \cite{cc-10} predicting the same behavior for any number of intervals. 
Thus as a first step we check the corrections to the scaling exponent for the Ising model in the absence of 
the Jordan-Wigner string between the two blocks (i.e. we consider only the correlation matrix 
$\Gamma_1$, as done in Ref. \cite{ffip-08}). In this case, the results reported in Fig. \ref{corrIfer} give a  
compelling evidence that the leading corrections to the scaling are given by $\delta_\a=2/\a$ 
as in the single interval (data not reported here for $\a=3,4$ show that this remains true).
At this point it is natural that the Jordan-Wigner string produces another operator at the conical singularity,
that in the Ising model is the leading one. 
According to Ref. \cite{cc-10} all the corrections to the scaling should be of the form 
$\delta_\a=2x/\a$, thus taking the result $\delta_2=1/2$ for granted (see also below for
further evidences), we conclude that the Jordan-Wigner string introduces an operator with scaling 
dimension $x=1/2$. 
Such an operator in the continuum limit of the Ising model exists and it is the Majorana fermion,
that has exactly the same features of the Jordan-Wigner string (i.e. same symmetry and same 
non-local character). Such an operator is clearly not present in the single interval case. 
These considerations allow us to conclude that the leading corrections to the scaling for the double 
interval entanglement in the Ising model are described by the exponent
\be
\delta_\a=\frac1\a\,.
\ee

\begin{figure}
\includegraphics[width=\textwidth]{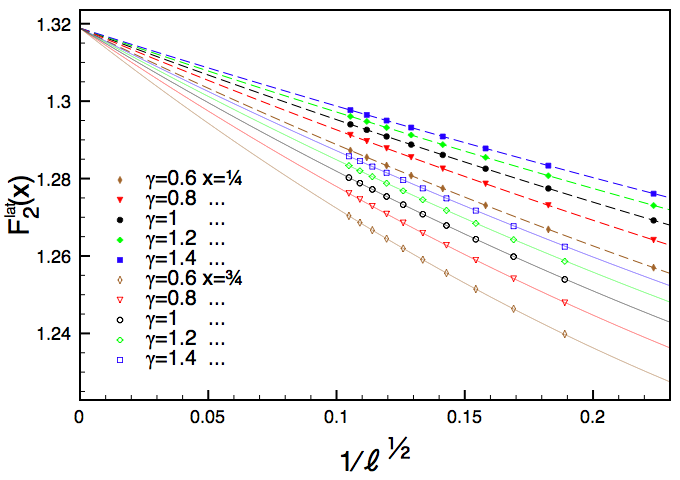}
\caption{Universality with respect to $\gamma$. We plot $F_2^{\rm lat}(x)$ for $x=1/4$ and $x=3/4$
for different values of $\gamma$ in the Hamiltonian. All results present the same leading correction to 
the scaling exponent $\delta_2=1/2$. The extrapolated data at $\ell\to\infty$ collapse in a single point 
equal to $F_2(1/4)=F_2(3/4)=1.31886\dots$. } 
\label{corrI}
\end{figure}

Unfortunately already for $\a=3$, the value of $\delta_3=1/3$ is very low and subleading corrections to 
the scaling going with exponents  $m\delta_\a$ (with $m$ integer) are expected to influence the results
in a considerable manner. 
For this reason, in order to have an accurate determination of the asymptotic 
behavior, at fixed $x$ we consider all corrections to the scaling up to those with exponent $1$. 
The resulting extrapolated data are the top points in Fig. \ref{ISvsCFT}.
In the case of $\a=2$, the CFT prediction $F_2(x)$ (cf. Eq. (\ref{CFTF2})).
 perfectly with the extrapolated data agree, giving strong support both for the procedure to account for 
the subleading corrections terms and on the  asymptotic form.  
For $\a\neq 2$ no CFT prediction is still available. 
The extrapolated data in Fig. \ref{ISvsCFT} are the first data for infinite $\ell$.

We now turn to consider the issue of universality. All the critical models ($h=1$) for any value 
of $\gamma\neq0$ are in the Ising universality class. However, the results at finite $\ell$ show a 
strong dependency on $\gamma$ (as obvious from the different correlation matrices).
In Fig. \ref{corrI} we report several data for $F_2^{\rm lat}(x)$ for different values of $\gamma$ and $\ell$
at fixed $x=1/4$ and $x=3/4$ (that for the symmetry $x\to1-x$ have the same asymptotic value).
At finite $\ell$, all results are evidently different. 
In the figure we report the extrapolation with two corrections to the scaling (i.e. with $\delta_2=1/2$ 
and $2\delta_2$).
For $\ell\to\infty$ all data tend to the same value predicted by CFT $F_2(1/4)=(2 + \sqrt2 (1 + 3^{1/4}))/4$, 
confirming in a single plot many results: (i) universality with respect to $\gamma$, 
(ii) correctness of the correction to the scaling form, 
(iii) correctness of the CFT prediction Eq. (\ref{CFTF2}).

\begin{figure}
\includegraphics[width=\textwidth]{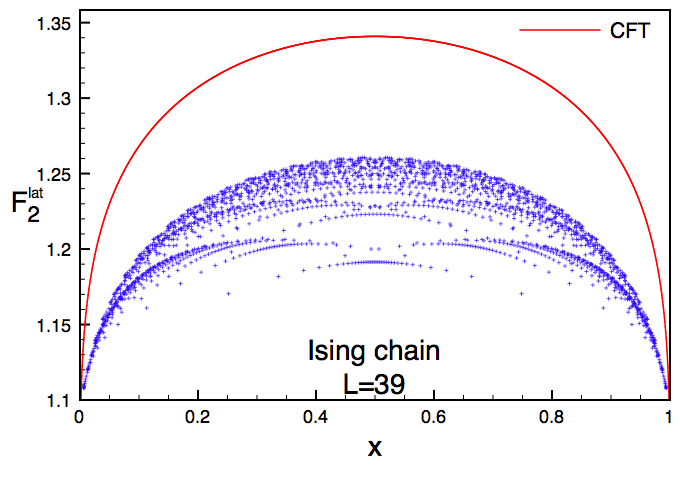}
\caption{The function $F_2^{\rm lat}(x)$ for a small chain with $L=39$ spins. 
Corrections to the scaling are monotonous and very large compared to the XX case: The data 
lie much below the asymptotic value predicted by CFT.} 
\label{caosIS}
\end{figure}

To conclude this section we report the data for a finite chain. As usual, in all the scaling variables 
we substitute distances with the chord distances. In Fig. \ref{caosIS}, we report the values of 
$F_2^{\rm lat}(x)$ for all the possible choices of intervals $A_{1,2}$ 
and $B_{1,2}$ in a chain of $L=39$ spins. 
Compared to the analogous plot for the XX chain (Fig. \ref{caosXX}), the figure is much clearer, due to the 
fact that corrections to the scaling are monotonous. 
Notice however that the data points lie much below the asymptotic value because the 
exponent $\delta_2=1/2$ is  small (also compared to the XX case with $\delta_2=1$)
resulting in very large corrections to the scaling.

\section{Non critical models}
\label{NCsec}

\begin{figure}
\includegraphics[width=0.55\textwidth]{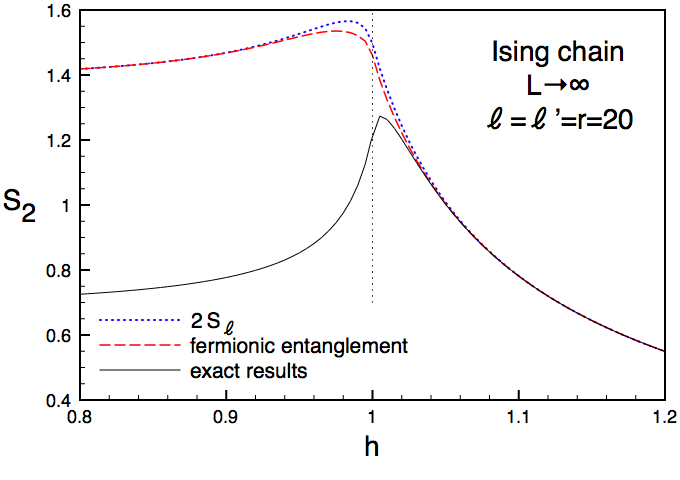}
\includegraphics[width=0.55\textwidth]{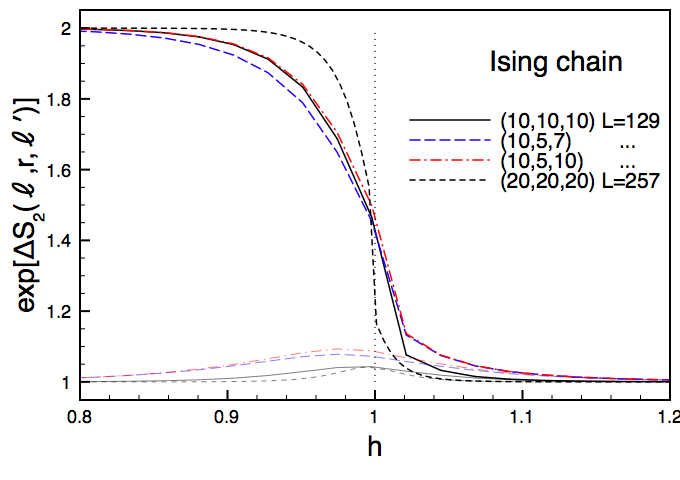}
\caption{Left: Comparison of the R\`enyi entropy $S_2$ for a non-critical Ising model 
of two disjoint intervals versus the double of the single interval and the fermionic one. 
While the fermionic entropy is asymptotically always the double of the single interval case, 
the one in spin variables presents a $-\log2$ difference due to the string. 
Right: Mutual entropy $\Delta S_2$ as a function of the magnetic field $h$ for an Ising chain of $129$
and $257$ spins. The crossing at the phase transition 
 is shown for three configurations. The intercept with $h=1$ is in good agreement with the CFT prediction. 
 The opaque lines are the corresponding fermionic mutual entropies} 
\label{NC}
\end{figure}

The richness of the phase diagram of the XY model allows us also to explore gapped phases that are 
almost everywhere except on the line $|h|=1$ and the segment $\gamma=0$ with $h^2<1$.
These systems have not be considered so much in the literature until now, because for 
non-critical models all the correlations and entanglement between the two blocks fall off exponentially 
(with a decay rate given by the inverse gap or mass).
Thus one would always expect
\be
 S_\a(\ell,r,\ell')= S_\a(\ell)+S_a(\ell')+O(e^{-r/\Delta})\,.
\ee
However, this is not so obvious because of the importance of the connected part in the 
correlations. In Fig. \ref{NC} (left) we report $S_2$ for the double interval case. 
While for $h>1$, the spin entropy is the same as the double of the single interval and the same 
as the fermionic one, for $h<1$ there is clearly an offset (that we quantify in $-\log2$, see below). 
The best way of detecting these unexpected effects is to consider the mutual entropy
\be
\Delta S_\alpha(\ell,r,\ell')=S_\alpha(\ell)+S_{\alpha}(\ell')-S_\alpha(\ell,r,\ell')\,,
\ee
that gives automatically zero when factorization occurs.

From numerical data we deduce that, for large blocks, all terms
\be
\prod_{i=1}^\alpha c[\zeta_i]\{\Gamma_{\zeta_1},\cdots,\Gamma_{\zeta_\alpha}\}\,,
\ee
in Eq. (\ref{eq:Salpha}) have the same absolute value. However,  they can have different signs. 
Furthermore numerical evidence suggests that terms consisting only of the correlation matrices $\Gamma_1$ and 
$\Gamma_2$ have always positive sign. 
If these observations are generally true, we have
\be
\Delta S_\a\sim \frac{1}{\a-1}\log\Bigl[1+\frac{1}{2^\a}\sum_{\{\zeta\} \ s.t. \atop{\#_3+\#_4\neq 0}}(-1)^{\#_4}\epsilon[\{\zeta\}]\Bigr]\,,
\ee
where \(\epsilon[\{\zeta\}]\) is the sign associated to the element \(\{\Gamma_{\zeta_1},\cdots,\Gamma_{\zeta_\alpha}\}\) and \(\#_{3(4)}\) is the number of  correlation matrices \(\Gamma_{3(4)}\).
We expect an eventual discontinuity in $\Delta S_\a$ when crossing a critical line. 
Thus we study non-critical chains with magnetic field close to $h=1$. 
We found numerically that only terms with odd number of correlation matrices $\Gamma_{3(4)}$ 
display sign changes. 
Thus (see right panel of Fig. \ref{NC} for the explicit plot) we conclude from the numerical evidence,  
the behavior
\be
\Delta S_\alpha=\cases{
\log 2 & $h^2<1$,\\
0&otherwise .
}
\ee
This result is not completely unexpected: 
also $S_A$ of the single interval for $|h|=0$ tends to $\log 2$, independently on 
$\ell$ \cite{cc-04,jk-04,p-04}. 
Thus in the definition of $\Delta S_\a$, since also the double interval $S_\a$ tends to the same value, 
we are left with a single $\log 2$.\footnote{This is a consequence of the double degeneration of the 
ground state for $|h|<1$. It is easily understood at $h=0$, where the ground state is any linear combination
of the states all up and all down that we can denote with $|\uparrow\rangle$ and $|\downarrow\rangle$.
Since there is no symmetry breaking term in the Hamiltonian, the diagonalization selects a state with 
zero magnetization, i.e. $(|\uparrow\rangle\pm |\downarrow\rangle)/\sqrt2$, in which the entanglement of 
any subsystem, connected or not, is always $\log2$ as stated in the main text.}

We stress that at $h=1$  the various $\Delta_\a$ cross in a single point if they are characterized 
by the same four point ratio $x$ (see right panel of Fig. \ref{NC}) that is the CFT prediction. 
These kind of plots could be used to detect the phase transition points in systems where it is not 
exactly known.

The $\a$ independence of the previous expression,  allows us to analytical continue the result to $\a=1$ 
and to conjecture the same behavior for the asymptotic von Neumann 
entanglement entropy in non-critical regions
\be
S_{1}
=S_1^{\rm ferm}
- \theta(1-h^2) \ln 2+\dots\,, 
\ee
where $S_1^{\rm ferm}$ is the fermionic entanglement entropy obtained from the correlation 
matrix $\Gamma_1$. Notice that this does not exclude that exponential corrections to this asymptotic 
form in the fermionic and spin variables could have different amplitudes. It would be interesting to explore 
this issue with the continuum theory in the form-factor approach \cite{ccd-07}.

\section{Entanglement evolution following a quench}
\label{quesec}

Finally we consider the evolution of  R\`enyi entropies  after global quenches. 
In this problem, the system is prepared in the ground-state $|\psi_0\rangle$ of an 
Hamiltonian $H_0$ of the form 
(\ref{HXY}) and then it is let evolve from time $t=0$ with a different Hamiltonian $H$,
always of the form (\ref{HXY}) but with $(h,\gamma)\neq (h_0,\gamma_0)$.
The main feature of this non-equilibrium problem is that the initial state differs globally from the ground 
state and the excess of energy (compared to the ground-state of $H$) is extensive.
A connected spin block reacts to the quench increasing  R\`enyi entropies linearly in time up to each spin 
in the subsystem becomes entangled with the environment. 
Then entropies saturate. 
There is no reason why disjoint blocks should behave in different way, except for the necessity to 
subtract the mutual entropy.  
Our only goal here is to understand the differences between the spin representation and the 
fermionic one, postponing any more accurate analysis to further studies. 
We only report results for the R\`enyi entropy $S_2$, but the value of $\a$ is unimportant.
We compare the numerical data with the prediction that follows from the interpretation of the 
entanglement evolution in terms of motion of quasiparticles \cite{cc-05}. 
According to this physical scenario, because the excess of energy of the system is extensive, 
any site acts as a source of quasiparticle excitations. 
Particles emitted from different points (further apart than the correlation length in the initial state) are 
incoherent, but pairs of particles moving to the left or right from a given point are entangled. 
Thus $S_\a(t)$ should just be proportional to the number of coherent particles that emitted from any 
point reach one a point in $A$ and the other a point in $B$. Since there is a maximum speed for these excitations $v_M$, for a single interval this implies the linear growth for $2v_M t < \ell$ and saturation 
for very large times. For a general bipartition the result is \cite{cc-05} 
\be\fl
S_\a(t)\approx \int_{x'\in A}dx'\int_{x''\in B}dx''\int_{-\infty}^\infty
dx\int
H_\a(p) dp\delta\big(x'-x-v(p)t\big)\delta\big(x''-x-v(p)t\big)\,,
\ee
where we assumed momentum conservation.
For a double interval, this formula predicts a series of linear behavior with different
slopes, that finally saturates at late time \cite{cc-05}.
The function $H_\a$ does not depend on the subsystem length and topology. 
It has been exactly derived for a single interval \cite{fc-08}
\be
H_\a(\cos\Delta)=\frac{1}{1-\a}\log\Bigl[\Bigl(\frac{1+\cos\Delta}{2}\Bigr)^\a+\Bigl(\frac{1-\cos\Delta}{2}\Bigr)^\a\Bigr]\,,
\ee
where $\Delta (p)$ is the difference between the Bogolioubov angles before and after the quench, whose 
explicit expression in term of the dispersion relation $\varepsilon(\varphi)$ is reported in  \cite{fc-08}. 
Thus, plugging everything together, the  leading order of R\`enyi entropies is 
\begin{equation}\label{eq:timeprediction}
S_{\ell_1,r,\ell_2}(t)\sim S_{\ell_1}(t)+S_{\ell_2}(t)-\Delta S[\ell_1,r,\ell_2](t)\, ,
\end{equation}
with (\(\chi_{[a,b]}(x)\equiv\theta(x-a)\theta(b-x)\))
\[\fl
\Delta S
(t)=\int_{-\pi}^\pi\frac{\mathrm{d}\varphi}{2\pi} H_\a(\cos\Delta)\int_r^{2|\varepsilon'_\varphi|t}\mathrm{d}v\Bigl(\chi_{[r,r+\min(\ell_1,\ell_2)]}(v)-\chi_{[r+\max(\ell_1,\ell_2),r+\ell_1+\ell_2]}(v)\Bigr)
\]
and \cite{fc-08}
\be
S_\ell(t)=t\int_{2|\varepsilon'|t<\ell}\frac{\mathrm{d}\varphi}{2\pi}2|\varepsilon'| H_\a(\cos\Delta)+\ell\int_{2|\varepsilon'|t>\ell}\frac{\mathrm{d}\varphi}{2\pi} H_\a(\cos\Delta)\, .
\ee

\begin{figure}
\includegraphics[width=\textwidth]{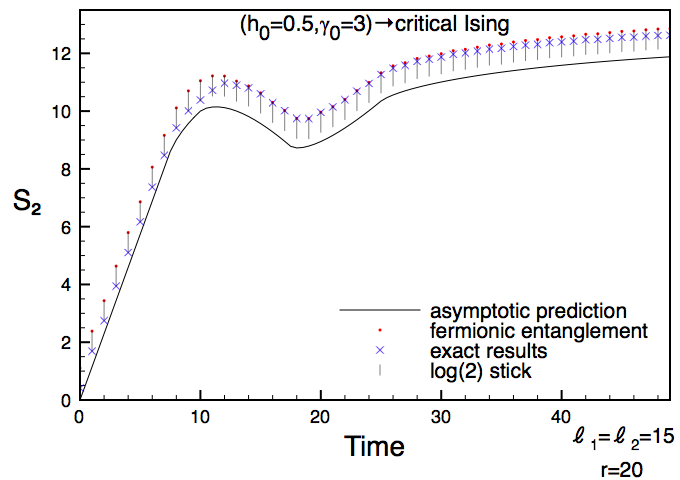}
\caption{Evolution of the R\`enyi entropy $S_2$ after a global quench from a non critical system to a 
critical Ising. The asymptotic prediction is given by equation (\ref{eq:timeprediction}) 
while the fermionic entanglement is obtained neglecting the string contribution. 
When $\{\Gamma_1,\Gamma_2\}$ becomes negligible with respect to $\{\Gamma_1^2\}$, 
a jump of about $\sim\log 2$ is seen. 
In the graph this happen in the neighborhood of the time $t\sim 11$. 
The asymptotic prediction presents a small offset due to the finite entanglement in the initial state.} 
\label{QQ}
\end{figure}

Let us now go back to the direct computation of the entanglement in two blocks after a quench, we observe 
that not all terms contribute to the leading order in blocks lengths and distance: 
the string expectation value decays exponentially and  only terms constructed 
with $\Gamma_1$ and $\Gamma_2$ survive.
For instance the Renyi entropy $S_2$ is simply the logarithm of two terms
\bea\fl
S_2&\sim&-\log\Bigl[\frac{1}{2}\exp\Bigl(\frac{1}{2}
\Tr\log \frac{1+\Gamma_1^2}{2}\Bigr)+\frac{1}{2}\exp\Bigl(\frac{1}{2}\Tr{\log \frac{1+\Gamma_1\Gamma_2}{2}}\Bigr)\Bigr]\sim \\ \fl
&\sim& \quad\cases{
-\frac{1}{2}\Tr{\log \frac{1+\Gamma_1^2}{2}}+\log{2}+c&if\, $\Tr{\log\frac{1+\Gamma_1\Gamma_2}{2}}<\Tr{\log\frac{1+\Gamma_1^2}{2}}$\\
-\frac{1}{2}\Tr{\log \frac{1+\Gamma_1^2}{2}}+c&if\, $\Tr{\log\frac{1+\Gamma_1\Gamma_2}{2}}\sim\Tr{\log\frac{1+\Gamma_1^2}{2}}$\\
-\frac{1}{2}\Tr{\log \frac{1+\Gamma_1\Gamma_2}{2}}+c+\log2& otherwise.
}
\eea
Analyzing several data (we report only a single example in Fig. \ref{QQ}), we deduce that in the scaling limit 
the term constructed with the standard fermionic correlations is never negligible: 
finite order Renyi entropies are controlled by fermionic correlations. 
When other terms are comparable with $\{\Gamma_1,\cdots,\Gamma_1\}$, they give additive $O(1)$
contribution of the form 
$\sim \log k$, where $k$ is the coefficient in front of the factor $\{\cdots\}$, and $k=2$ for $S_2$. 

The quasi-particle interpretation \cite{cc-05} perfectly agrees (up to a constant) with the 
fermionic representation $\rho_{\Gamma_1}$ (see Fig. \ref{QQ}). 
In other words the non locality of the Jordan-Wigner transformation does not influence (as one expects) 
the leading order of  time evolution of entropies after a global quench. 
However, as in the non-critical equilibrium case, we observe an additive $\log 2$ contribution 
(in Fig. \ref{QQ} a stick of width $\log2$ is included to appreciate this difference)  from the 
string and we believe that this is a general feature that will persist also away from criticality. 
As a final comment, we mention that the $O(1)$ offset of the numerical data at finite $\ell$ compared with 
the asymptotic form is a consequence of the initial entanglement. This subleading term is present 
also for the single interval case \cite{fc-08}, and can be included in the asymptotic result \cite{sc-08}.

\section{Discussions}
\label{disc}

We presented a general method to calculate R\`enyi entropies for integer $\a$ in the ground state
of the XY Hamiltonian (\ref{HXY}) in the case when $A$ consists of two (but the method works for even 
more) disjoint intervals.
We carefully analyze the  results in the critical cases of $XX$ and Ising universality classes. 
We found that the asymptotic results for large intervals and separations are described correctly by conformal
field theory when results are known. 
We also provided results for the asymptotic scaling of the $\a=3,4$ R\`enyi entropies of 
the Ising model, for which no analytic prediction is still available.
In order to check CFT predictions we had to carefully include corrections to the scaling in the analysis. 
We found that for the XX model, the leading corrections to the scaling are governed by the 
exponent $\delta_\a=2/\a$, the same as the single interval case \cite{ccen-10}, in agreement 
with general predictions \cite{cc-10}. 
Oppositely for the critical Ising model we found $\delta_\a=1/\a$, that is half of the single interval one. 
This can be understood in terms of the general theory of the corrections to the scaling \cite{cc-10}
by interpreting the Jordan-Wigner string in the lattice model as a non-local fermion in the Ising CFT.
We also have briefly considered the entanglement in the gapped phase and the time evolution 
following a quantum quench. The provided exact representation of the reduced density matrix of 
spin variables could also be useful for the calculation of the entanglement between 
$A_1$ and $A_2$ measured by the negativity \cite{Neg}.

A major problem remains still open. We are unable to find all the eigenvalues of the reduced density 
matrix of two intervals and so the R\`enyi entropies for non-integer $\a$ and in particular for $\a=1$ 
that would give the widely studied von Neumann entropy. The same 
problem is also present in conformal calculations. It is possible to obtain $\Tr\rho_A^\a$ only for 
$\a$ integer \cite{cct-09} and the analytic continuation of the result to general complex values remains 
a big open problem. 

The contribution of the Jordan-Wigner string also affects the entanglement entropy of 
a single interval in a system with one or more boundaries, if the block $A$ does not include the 
boundary. In this case, the correlation matrix is slightly more complicated and we are currently 
analyzing the problem. Among the other things, this would be relevant also for some quench problems
where indeed only the fermionic entanglement has been calculated \cite{ep}.

Another interesting open question concerns the behavior of the R\'enyi entropies in systems 
with disorder for which in the single block case it is known that the entanglement simply follows from 
the counting of the singlets \cite{rm-05} between A and B. Whether this remains true in the case of more 
intervals is not yet known.

\appendix

\section{Correlation matrices} \label{Gamma3}

We report in this appendix some simple properties of the correlation matrices
\be
\Gamma_{i j}^S=\Tr[a_i\rho^S a_j]-\delta_{i j}\qquad i,j\in A\, ,
\ee
with $\rho_S$ defined in Eq. (\ref{rhoSA}).
$\Gamma^S$'s elements are ratio of two expectation values ($n\neq0$)
\begin{equation}\label{eq:APPGamma}
 \Gamma_{l l+n}^S=\frac{\Bigl<a_{l+n}\prod_{j\in B_1}(a_{2j-1}a_{2j}) a_{l}\Bigr>}{\Bigl<\prod_{j\in B_1}(a_{2j-1}a_{2j})\Bigr>}\,,
\end{equation}
which can be evaluated by means of the Wick theorem. 
We show that $\Gamma^S$ can be expressed in terms of the system correlation matrix 
$\Gamma\equiv \Gamma^{\bf 1}$, where $\bf 1$ is the identity matrix. 
By isolating the two-point correlations between $a_l$ and the other 
Majorana operators in the numerator of (\ref{eq:APPGamma}) we obtain 
\be
\Bigl<a_{l+n}\prod\limits_{j\in B_1}(a_{2j-1}a_{2j}) a_l\Bigr>=
\langle S\rangle\Gamma_{l\, l+n}+\bigl[\Gamma \Lambda \Gamma\bigr]_{l\, l+n} \, ,
\ee
with
\be
\Lambda_{i j}=(-1)^{i-j} {\rm sign}(j-i)\Bigl<\prod_{k\in B^*_{ij}} a_{k} \Bigr>\, ,
\label{eq:Lambda}
\ee
where $B^*_{ij}$ denotes the set of all indices $ \{2l\}\cup\{2l-1\}$ with  $l\in B_1$ and with $i$ and $j$ 
removed.  
The expectation value in (\ref{eq:Lambda})  is the Pfaffian of the system correlation matrix restricted to 
the region $B^*_{ij}$.
Checking the Pfaffian sign we have
\be
\Lambda_{i j}=-\langle S\rangle[\Gamma_{B_1}^{-1}]_{i j}\, ,
\ee
with $\Gamma_{B_1}$ the correlation matrix restricted to the region $B_1$.
Substituting this  expression into (\ref{eq:APPGamma}) we get
\be
\Gamma^S=\Gamma_A-\Gamma_{A B_1}\frac{1}{\Gamma_{B_1}}\Gamma_{B_1 A}
\equiv \Gamma_{A\cup B_1}/\Gamma_{B_1}\, ,
\ee
where the double subscript takes into account restrictions to rectangular correlation matrices, i.e. 
the first (second) subscript identifies the region where the row (column) index runs. 
Notice that $\Gamma^S$ is the Schur complement of $\Gamma_{B_1}$ in $\Gamma_{A\cup B_{1}}$.
This proves Eq. (\ref{sh}).

\section{Generalization to many disjoint spin blocks}
\label{moreint}

We consider in the appendix how to generalize our approach to the case when the subsystem consists 
of $n$ disjoint blocks of lengths $\ell_i$, with $i=1,\dots,n$. 
We indicate with $O_i^{+}$ a product of an even number of  Majorana operators lying in the $i$-th 
block $A_i$ and with $O^{-}_i$ an odd product. 
$S_i^B$ is the complete $\sigma_z$ string associated to the $i$-th region between consecutive 
blocks $B_i$ and $S_i^A$ is the string corresponding to the block $A_i$ (see Fig. \ref{chain} for the double 
block case). Eq. (\ref{eq:rhofer}) can be easily generalized observing that the reduced density matrix 
has the following  expansion
\be\fl
\rho_{\cup_i A_i}=\frac{1}{2^{\sum_i \ell_i}}
\sum_{\{s\}}\sum_{O_1^{s_1}\cdots O_n^{s_n}}\langle S[\{s\}] O_1^{s_1}\cdots O_n^{s_n}\rangle
(O_1^{s_1}\cdots O_n^{s_n})^\dag S[\{s\}]\, ,
\ee
where $s_i=\pm1$ and
$S[\{s\}]$ are $\s_z$ strings determined by the conditions
\bea\label{eq:APPconditions}
S[\{s_1,\cdots,s_k,\cdots s_n\}]&=&S[\{s_1,\cdots,-s_k,\cdots s_n\}]\prod_{i=1}^{k-1}S_i^B\,,\\ \nonumber
S[\{0,\cdots,0\}]&=&1\, .
\eea
Indeed, any spin operator that can be written as a product of Majorana operators that are odd in number 
if restricted to the $k$-th block is constructed with an odd number of $\sigma^{x(y)}$ of the $k$-th block. 
Spin variables $\s^x$ and $\s^y$ are non-local in the fermionic space
\be
\s^{x(y)}_{l\in A_k}=\Lambda (A_1,\dots,A_{k-1})\prod_{i=1}^{k-1}S_i\ a_{2l(2l-1)}\, ,
\ee
where $\Lambda(A_1,\dots,A_{k-1})$ is an operator that depends on the Majorana fermions of the 
blocks $A_1,\dots A_{k-1}$. When multiplying an odd number of $\s^{x(y)}$ the string 
$\prod_{i=1}^{k-1}S_i$ does not cancel (we remind $S_i^2=1$) so we obtain the
conditions (\ref{eq:APPconditions}).
The construction of the fermionic equivalent reduced density matrix is analogous to the double block case. In fact the unitary transformation 
\[
U=\sum_{\{s\}_n}\prod_{i=1}^n\Pi_i^{s_i}\qquad s_i\in\{+,-\}\, ,
\]
with
\be
\Pi_i^+=\frac{1+S_i^A}{2}\,,\qquad
\Pi_i^-=\prod_{j=1}^{i-1}S_j^B\frac{1-S_i^A}{2}\,,
\ee
brings the spin reduced density matrix to a fermionic one. In particular we obtain the simple result 
\be
\rho_{\cup_iA_i}\sim \frac{1}{2^{\sum_i \ell_i}}
\sum_{\{s\}}\sum_{O_1^{s_1}\cdots O_n^{s_n}}\langle S[\{s\}] O_1^{s_1}\cdots O_n^{s_n}\rangle(O_1^{s_1}\cdots O_n^{s_n})^\dag\, . 
\ee
The term associated to the configuration $\{s\}$
\be
\rho_{\{s\}}^{S[\{s\}]}=\frac{1}{2^{\sum_i \ell_i}}
\sum_{O_1^{s_1}\cdots O_n^{s_n}}\langle S[\{s\}] O_1^{s_1}\cdots O_n^{s_n}\rangle(O_1^{s_1}\cdots O_n^{s_n})^\dag
\ee
is sum of $2^{n-1}$ fermionic RDMs.
This is because, fixed $S[\{s\}]$, the further dependence on $s_i$ can be handle in the following way 
(remind $S_k^A O_k^s S_k^A= s O_k^s$)
\bea
\rho_{\{s_1,\cdots,s_k,\cdots, s_n\}}^{S[\{s\}]}&=&
\frac{[\rho_{\{s_1,\cdots,s_k,\cdots, s_n\}}^{S[\{s\}]}+\rho_{\{s_1,\cdots,-s_k,\cdots, s_n\}}^{S[\{s\}]}]
}{2}+\nonumber\\ &&+
\frac{S_k^A[\rho_{\{s_1,\cdots,s_k,\cdots, s_n\}}^{S[\{s\}]}+\rho_{\{s_1,\cdots,-s_k,\cdots, s_n\}}^{S[\{s\}]}]S_k^A}{2}\,,
\eea
and, for any string \(S^B\) and \(S^A\), the operator
\be
S^A\sum_{\{s\}}\rho^{S^B}_{\{s\}}S^A\,,
\ee
is a fermionic RDM. 
We obtain $2^{n-1}$ and not $2^n$ fermionic RDMs only because $S[\{s\}]$ does not depend 
on $s_1$ (there is no string contribution for operators lying in the first block).
Thus we can write 
\begin{equation}\label{eq:rhonblocks}\fl
\rho_{\cup_i A_i}\sim\frac{1}{2^{n-1}}\sum_{\{i\}{\rm w/o\, repetition}}
{\rm Pf} \Bigl[\Gamma_{\bigcup_{j\in\{i\}}B_j}\Bigr]\sum_{\{s\}}\epsilon_{\{i\}}^{\{s\}}\rho[\Gamma_{\{i\}}^{\{s\}}]\qquad s_i\in \{-1, 1\}\, .
\end{equation}
The explicit form of the correlation matrices \(\Gamma_{\{i\}}^{\{\sigma\}}\) is a simple generalization of the double block case
\be\fl
\Gamma_{\{i\}}^{\{s\}}=P[s_1,\dots,s_{n-1}]\Gamma_{A \cup(\bigcup_{j\in\{i\}}B_i) }/\Gamma_{\bigcup_{j\in\{i\}}B_j}P[s_1,\dots,s_{n-1}]\, ,
\ee
where we indicate with $A/B$ the Schur complement of \(B\) in \(A\) (see \ref{Gamma3}). 
${\rm Pf}$ denotes the Pfaffian, and $P$ is the diagonal matrix
\be\fl
P[s_1,\dots, s_{n-1}]=\left(
\begin{array}{ccccc}
s_1 1_{A_1}&&&&\\
&s_2 1_{A_2}&&&\\
&&\ddots&&\\
&&&s_{n-1} 1_{ A_{n-1}}&\\
&&&&1_{A_{n}}
\end{array}\right)\, .
\ee
The index $i\in\{0,n-1\}$ runs over the environment blocks lying between the first and the last 
subsystem block, and $B_0\equiv \emptyset$ so that \(\Gamma_{\{0\}}\) is the standard fermionic 
correlation matrix \(\Gamma_{A}\equiv\Gamma_{\bigcup_i A_i}\).
The sign \(\epsilon\) is determined by the conditions
\be\fl
\epsilon_{\{i\}}^{\{s\}}=\cases{
1&$\{i\}=\{0\}\vee\{s\}=\{1,\dots,1\}$\,,\\
(-1)^{j\in\{i\}} \epsilon_{\{i\}}^{\{s\}_{s_j\to-s_j}}&otherwise\, ,
}
\ee
where \((-1)^{j\in\{i\}}\) is equal to \(-1\) if \(j\in\{i\}\) and \(1\) otherwise, and
\be
\{s\}_{s_j\to -s_j}=\{s_1,\cdots,s_{j-1},-s_j,s_{j+1},\cdots,s_{n-1}\}\, .
\ee
R\`enyi entropies can be computed in the same way as when the subsystem consists of just two disjoint 
blocks. But many more terms contribute. And when the number of blocks is comparable with the chain 
size we expect an extensive behavior of Renyi entropies, as observed in \cite{ip-09}, that is 
reminiscence of the huge number of fermionic RDMs needed to represent the spin RDM. 
Finally, as in the double block case, R\`enyi entropies do not depend on the sign of Pfaffians because 
any sign change is the result of a unitary transformation (that is just a $\sigma_z$ string).

\section{Some technical details for the product rules}\label{app:Skew-symmetry}

The reduced density matrix of a given number of disjoint spin blocks in a free chain is unitarily equivalent
 to a finite sum of operators that well behave under multiplication. 
 In fact one of our results is that, for some $\sigma_z$ strings $S_j$ and constants $C_j$, the relation
\be
\rho\sim\sum_jC_j\frac{\Tr_{\bar A}{S_j\rho_0}}{\langle S_j\rangle }\equiv\sum_j C_j \rho^{S_j}\,,
\ee
holds. 
Like in the single block case  each operator $\rho^{S_j}$ turns out to be a normalized exponential of a quadratic form in the Majorana fermions.
Up to a constant factor,  these matrices are closed under multiplication and we identify 
any of them  by the skew-symmetric matrix $W$ in the exponent in Eq. (\ref{eq:rhoW}).
We recall in this appendix some basic properties of skew-symmetric diagonalizable matrices.

If $\lambda$ is an eigenvalue of $W$, then also $-\lambda$ is because 
\be
0=\det\bigl|W-\lambda I\bigr|=\det\bigl|W^T-\lambda I\bigr|=\det\bigl|-W-\lambda I\bigr|\, .
\ee

If $\vec w_\lambda$ and $\vec w_\mu$ are two eigenvectors  with eigenvalues $\lambda$ and $\mu$, then
\be\fl
\mu \vec w_\lambda\cdot \vec w_\mu=\vec w_\lambda^T W \vec w_\mu=-\vec w_\mu^T W \vec w_\lambda=-\lambda \vec w_\mu\cdot \vec w_\lambda\, ,
\quad \Rightarrow \quad
\label{eq:APPortho}
(\mu+\lambda)\vec w_\mu\cdot\vec w_\lambda=0\, .
\end{equation}
The vectors \(\vec w_\mu\) and \(\vec w_\lambda\) are orthogonal (with respect to the 
scalar product  $\vec w_\mu\cdot\vec w_\lambda =\sum_j (w_\mu)_j (w_\mu)_j$, that is not the standard one 
because of the absence of complex conjugation) unless \(\mu=-\lambda\).
If the eigenvalue \(\lambda\) (and so \(-\lambda\)) is \(n\)-times degenerate, in general one can redefine 
the eigenvectors in the same \(n\)-dimensional eigenspace so that each vector \(\vec w_\lambda^i\) 
becomes orthogonal to the other $n-1$ ones.
Thus the following \emph{decomposition} holds
\begin{equation}\label{eq:APPdecomposition}
W=\sum_{\{\lambda\}/_{\pm}\atop \vec w_\lambda\cdot \vec w_{-\lambda}\neq0}\lambda \frac{\vec w_\lambda\otimes\vec w_{-\lambda}-\vec w_{-\lambda}\otimes\vec w_{\lambda}}{\vec w_\lambda\cdot \vec w_{-\lambda}}\equiv \sum_{\{\lambda\}/_\pm}\lambda \ \Pi_\lambda^{-}\, ,
\end{equation}
where \(\{\lambda\}/_\pm\) means that each couple \(\pm \lambda\) contributes only once in the sum and the symbol \(\vec v \otimes \vec w\) identifies the matrix with elements \(v_i w_j\). 
By introducing the auxiliary matrices
\be
\Pi_\lambda^+\equiv \frac{\vec w_\lambda\otimes\vec w_{-\lambda}+\vec w_{-\lambda}\otimes\vec w_{\lambda}}{\vec w_\lambda\cdot \vec w_{-\lambda}}\, ,
\ee
which  satisfy  together with \(\Pi_\lambda^-\) the orthogonality relations
\be
\begin{array}{c|cc}
\cdot&\Pi_\lambda^+&\Pi_\lambda^-\\
\hline 
\Pi_\mu^+&\delta_{\mu+\lambda}\Pi_\mu^+&\delta_{\mu+\lambda}\Pi_\mu^-\\
\Pi_\mu^-&\delta_{\mu+\lambda}\Pi^-_\mu&\delta_{\mu+\lambda}\Pi_\mu^+\, ,
\end{array}
\ee
any function of \(W\) expresses in terms of \(\Pi^\pm_\lambda\)
\be\fl
f(W)=\sum_{\{\lambda\}/_\pm} \Bigl(f_e(\lambda) \Pi^+_\lambda+f_o(\lambda) \Pi^-_\lambda\Bigr)=\sum_{\{\lambda\}/_\pm} \frac{f(\lambda) \vec w_\lambda \otimes\vec w_{-\lambda}+f(-\lambda) \vec w_{-\lambda} \otimes\vec w_{\lambda}}{\vec w_\lambda \cdot \vec w_{-\lambda}}\, ,
\ee
with \(f_e\) and \(f_o\) the even and odd part of \(f\).

The decomposition (\ref{eq:APPdecomposition}) allows to express $\rho_W$ in a factorized form 
\bea\fl 
Z(W)\rho_W&=&
\exp\Bigl(\sum_{\{\lambda\}/_{\pm}}\frac{\lambda}{2} \frac{[\vec w_\lambda\cdot \vec a,\vec w_{-\lambda}\cdot \vec a]}{2 \vec w_\lambda\cdot \vec w_{-\lambda}}\Bigr)=\prod_{\{\lambda\}/\pm}\exp\Bigl(\frac{\lambda}{2} \frac{[\vec w_\lambda\cdot \vec a,\vec w_{-\lambda}\cdot \vec a]}{2 \vec w_\lambda\cdot \vec w_{-\lambda}}\Bigr)= \nonumber\\
\fl&=&\prod_{\{\lambda\}/\pm}\Bigl[\cosh\Bigl(\frac{\lambda}{2}\Bigr)+\sinh\Bigl(\frac{\lambda}{2}\Bigr)\frac{[\vec w_\lambda\cdot a,\vec w_{-\lambda}\cdot a]}{2 \vec w_\lambda\cdot \vec w_{-\lambda}}\Bigr]\, ,
\label{eq:APPrhoWfactor}
\eea
where we made use of \(\vec w_\lambda\)'s orthogonality (\ref{eq:APPortho})
\be\fl
\Bigl[[\vec w_\lambda\cdot \vec a,\vec w_{-\lambda}\cdot \vec a],[\vec w_\mu\cdot \vec a,\vec w_{-\mu}\cdot \vec a]\Bigr]=0\,,\qquad 
[\vec w_\lambda\cdot \vec a,\vec w_{-\lambda}\cdot \vec a]^2=4(\vec w_\lambda\cdot\vec w_{-\lambda})^2\, .
\ee
We obtain \(Z(W)\) by imposing the normalization of \(\rho_W\)
\be\fl
Z(W)= \Tr{\exp\Bigl(\frac{a W a}{4}\Bigr)}=\prod_{\{\lambda\}/_\pm}2 \cosh\Bigl(\frac{\lambda}{2}\Bigr)=\pm \sqrt{\det\bigl| e^{\frac{W}{2}}+e^{-\frac{W}{2}}\bigr|}\, .
\ee
The correlation matrix \(\Gamma_{i j}=\delta_{i j}-\bigl<a_ia_j\bigr>\) comes  directly from the quadratic part of expression (\ref{eq:APPrhoWfactor})
\bea\fl
Z(W)\Gamma_{i j}&=&\Tr{\frac{[a_j,a_i]}{2}\exp\Bigl(\frac{\vec a^T W \vec a}{4}\Bigr)}=\nonumber\\ \fl &=&
\prod_{\{\lambda\}/_\pm}\!\!\! \cosh\Bigl(\frac{\lambda}{2}\Bigr) 
\sum_{\{\lambda\}/_\pm} \tanh\Bigl(\frac{\lambda}{2}\Bigr)\Tr{\frac{[a_j,a_i]}{2}\frac{[\vec w_\lambda\cdot \vec a,\vec w_{-\lambda}\cdot \vec a]}{2 \vec w_\lambda\cdot \vec w_{-\lambda}}}= \nonumber\\ \fl
&=&\Bigl[\prod_{\{\lambda\}/_\pm}\!\!\! 2\cosh\Bigl(\frac{\lambda}{2}\Bigr)\Bigr]\tanh\Bigl(\frac{W}{2}\Bigr)\Bigr]_{ij}\, ,
\eea
that is
\be
\Gamma=\tanh\Bigl(\frac{W}{2}\Bigr)\, .
\ee
The correlation matrix does not determine univocally function \(Z(W)\). Indeed we obtain the expression
\begin{equation}\label{eq:APPZ(w)}
Z(W)\sim\frac{2^N}{\sqrt[4]{\det\bigl|1-\Gamma^2\bigr|}}\, ,
\end{equation}
with an evident phase ambiguity (that is actually a sign ambiguity).
In spite of the ambiguity in \(Z(W)\sim Z[\Gamma]\), we believe that the trace of two fermionic RDMs 
\be
\Tr[{\rho_W\rho_{W'}}]=\frac{Z(\log(e^W e^{W'}))}{Z(W)Z(W')}
\ee
is a functional of $\Gamma=\tanh(W/2)$ and $\Gamma'=\tanh(W'/2)$. In fact, 
we argue that 
\begin{equation}\label{eq:ProdRule}
\{\Gamma,\Gamma'\}=\Tr{\rho[\Gamma]\rho[\Gamma']}=\frac{1}{2^N}\prod_{\{\lambda\}/2}\lambda
\end{equation}
where the product is extended over all eigenvalues \(\lambda\) of \(1+\Gamma\Gamma'\) 
with degeneracy (always even) reduced by half.
We have not a rigorous proof of Eq. (\ref{eq:ProdRule}), but we give an argument to justify it.
After infinitesimal variations of   \(\Gamma\)  and \(\Gamma'\), function  \(Z(W)\) changes in the following way
\be
\mathrm d\log Z(W)=\frac{1}{2}\Tr{\frac{\Gamma\mathrm d \Gamma}{1-\Gamma^2}}\, ,
\ee
thus from equation (\ref{eq:APPZ(w)}) it follows
\be
\mathrm d \log \Tr{\rho_W\rho_{W'}}=\frac{1}{2}\Tr{\frac{\Gamma\times \Gamma'\mathrm d (\Gamma\times \Gamma')}{1-(\Gamma\times \Gamma')^2}-\frac{\Gamma\mathrm d \Gamma}{1-\Gamma^2}-\frac{\Gamma'\mathrm d \Gamma'}{1-{\Gamma'}^2}}\, .
\ee
Using Eq. (\ref{eq:GtimesGprime}) for $\Gamma\times\Gamma'$, many simplifications occur, leading to
\be\fl
\mathrm d\log \Tr[\rho_W\rho_{W'}]=\frac{1}{2}\Tr{\Gamma'\frac{1}{1+\Gamma\Gamma'}\mathrm d \Gamma+\frac{1}{1+\Gamma\Gamma'}\Gamma\mathrm d\Gamma'}
=\frac{1}{2}\Tr{\frac{\mathrm d (\Gamma \Gamma')}{1+\Gamma\Gamma'}}\, .
\ee
The variations \(\mathrm d \Gamma\) and \(\mathrm d \Gamma'\) are skew-symmetric by construction.
Our task is to integrate the differential equation above.
We consider all possible sets of \(2N\) smooth simple curves joining the \(\Gamma\Gamma'\) eigenvalues 
\(\lambda_i\) from  \(\Gamma\Gamma'=0\) (when \(\Tr{\rho_W\rho_{W'}}=2^{-N}\)) to our desired values. Despite of the initial ambiguity in the sign of \(Z(W)\), the expression for the logarithm
\be
\log \Tr[\rho_W\rho_{W'}]=
\frac{1}{2}\sum_{i=1}^{2N}\int_{\gamma_i}\frac{\mathrm d \lambda_i}{1+\lambda_i}-N\log 2\, .
\ee
is well defined, in view of the fact that \(\Gamma\Gamma'\) has a double-degenerate spectrum and the ambiguity related to the choice of the curves \(\gamma_i\) (avoiding any singular point) gives simply an additive constant proportional to \(2\pi i\).
Considering the path characterized by \(d(\Gamma\Gamma')=\mathrm d \lambda (\Gamma\Gamma'-1)/2\), the product rule can be expressed in the following integral form
\[\fl
\{\Gamma,\Gamma'\}=\exp\Bigl[\frac{1}{2}\int\limits_{\gamma_{[0\rightarrow 1]}}\Tr{\frac{\Gamma\Gamma'-1}{2-\lambda+\lambda\Gamma\Gamma'}}\mathrm d \lambda\Bigr]=\exp\Bigl[\frac{1}{2}\int\limits_{\gamma_{[0\rightarrow 1]}}\frac{1}{1+\lambda}\Tr{\frac{\Gamma\Gamma'-1}{1+\lambda\Gamma\Gamma'}}\mathrm d \lambda\Bigr]\, ,
\]
where \(\gamma\) is a smooth simple path in \(\mathbf C\) which joins \(0\) to
\(1\) and avoids any of the points at which \(1+\lambda\Gamma\Gamma'\) fails to be invertible. 
Because of the  factor $1/2$ in front of the integral we get equation (\ref{eq:ProdRule}).
For this proof we used several properties about Pfaffian reviewed in Ref. \cite{p-89}.
Summarizing 
\be
\rho[\Gamma]\rho[\Gamma']=\exp\Bigl(\frac{1}{2}\int_{\gamma_{[0\rightarrow 1]}}\frac{1}{1+\lambda}\Tr{\frac{\Gamma\Gamma'-1}{\lambda\Gamma\Gamma'+1}}\mathrm{d}\lambda\Bigr)\rho[\Gamma\times\Gamma']\, .
\ee

A final remark about the XY correlation matrices in the double block case \(\Gamma_1=\Gamma_{A}\) and \(\Gamma_3=\Gamma_{A\cup B_1}/\Gamma_{B_1}\) (\ref{Gamma3}). The quantity
\be
s\equiv\frac{\{\Gamma_1,\Gamma_3\}}{\sqrt{\det\bigl|\frac{1+\Gamma_1\Gamma_3}{2}\bigr|}}
\ee
can be \(1\) or \(-1\) and numerical data suggest that it is independent of the subsystem and of the chain length. 
In particular we find
\be
s=\cases{
1 &$|h|\geq 1\, {\rm or}\, \gamma=0$,\\
-1 &otherwise,
}
\ee
so that
\be
\{\Gamma_1,\Gamma_3\}=\Bigl[\mathrm{sign}(|h|-1)+2\delta_{\gamma 0}\theta(1-|h|)\Bigr]\sqrt{\det\Bigl|\frac{1+\Gamma_1\Gamma_3}{2}\Bigr|}\, .
\ee

\begin{figure}[t]
\begin{center}
\includegraphics[width=0.8\textwidth]{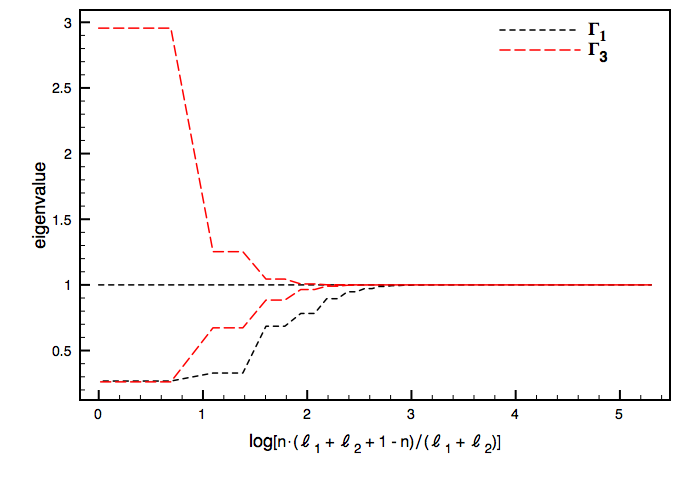}
\caption{Sorted absolute value of the eigenvalues of $\Gamma_1$ and $\Gamma_3$ 
for two identical blocks of length $\ell_1=\ell_2=100$ at the distance $r=50$ of an infinite Ising chain. 
$n$ is the eigenvalue's positioning number. The eigenvalues of $\Gamma_1$ lie between \(-1\) and \(1\) 
while the eigenvalues of \(\Gamma_3\) can be higher (lower) than \(1\) (\(-1\)). 
Almost all eigenvalues are in the neighborhood of \(1\) and \(-1\).}
\label{fig:spectrum}
\end{center}
\end{figure}

We close this appendix with some speculations and some simplified (wrong) expressions to which one 
could arrive if treating too naively correlation matrices. 
If we would not worry about the signs and non-invertibility,  we could write directly the result 
of a generic product like Eq. (\ref{boh}) as
\begin{equation}\label{eq:prodrulewrong}\fl
\{\Gamma_1,\cdots,\Gamma_n\}\sim \sqrt{\frac{\det\Bigl|\prod_{i=1}^n e^{ W_i}+1\Bigr|}{\prod_{i=1}^n\det\Bigl|e^{W_i}+1\Bigr|}}=\sqrt{\prod_{i=1}^n\det\Bigl|\frac{1-\Gamma_i}{2}\Bigr|\det\Bigl|\prod_{i=1}^n\frac{1+\Gamma_i}{1-\Gamma_i}+1\Bigr|}\, .
\end{equation}
This expression is generally wrong.
When substituting the chain correlation matrices,  $(1-\Gamma_{1,3})$ is usually non-invertible (Fig. 
\ref{fig:spectrum}). Thus, except for $S_2$, equation (\ref{eq:prodrulewrong}) is wrong. 

By using this wrong expression one can try to see if any simplification occurs in the general R\`enyi entropy.
In general it is not possible to push this approach, unless taking further wrong assumptions. 
Assuming for example the (wrong) hypothesis of commuting $\Gamma$ matrices, after trivial 
algebra we have (the wrong expression)
\be\fl
S_\a 
{\approx}\frac{1}{1-\alpha}\log\left[\frac{1}{2^\alpha}\sum_{\zeta_1\dots\zeta_{\alpha}}\prod_{i=1}^\alpha c[\zeta_i](\pm)\sqrt{\det\Bigl|\prod_{i=1}^\alpha\frac{1-\Gamma_{\zeta_i}}{2}+\prod_{i=1}^\alpha\frac{1+\Gamma_{\zeta_i}}{2}\Bigr|}\right]\,.
\ee
We wrote this expression here, because one could have wonder whether it (despite wrong) 
would have been defined for any real $\a$ 
(i.e. depending only on the eigenvalues of the matrices $\Gamma$) to have an expression analytically 
continuable to $\a\to1$ (and maybe correct in this limit). 
However, even this simplified wrong expression is only calculable for integer $\a$ and thus 
we did not insist in this direction and we preferred to exploit the exact solution without assumptions.

\section{Accidental Singularities}\label{app:Miscellanea}

When the correlation matrix of the region interposed between the blocks \(\Gamma_{B_1B_1}\) is 
non-invertible, that is the \(\sigma_z\) string mean value vanishes \(\langle S\rangle=0\), some problems arise. The density matrix (\ref{rhopm}) is not well defined and all successive discussions are meaningless. 
Here we give some tricks to solve these problems.
We consider a generic operator 
\be
Q_W=\frac{1}{2^{L}}\sum_{O\in A}\langle{e^{\frac{a W a}{4}} O}\rangle O\, .
\ee
If \(\langle{e^{\frac{a W a}{4}}}\rangle\neq 0\), as we observed earlier, \(Q_W\) is proportional to an effective free density matrix
\be
Q_W=\langle{e^{\frac{a W a}{4}}}\rangle\rho_{\Gamma[W]}\, .
\ee
But when \(\langle{e^{\frac{a W a}{4}}}\rangle\) vanishes (it can happen only when \(W\) is not hermitian) the previous expression is meaningless. 
The trick is to control the singularities introducing the traceless string operator \(S_E\equiv \prod_{j\in E}{ia_{2j-1}a_{2j}}\) (\(S_E^2=1\)) that belongs to the subsystem \(E\subset A \)
\be\fl
Q_W=S_E\frac{1}{2^L}\sum_{O\in A }\langle{e^{\frac{a^T W a}{4}} O}\rangle S_E O= 
S_E \frac{1}{2^\ell}\sum_{O\in A}\langle{e^{\frac{a W a}{4}} S_E O}\rangle O
\equiv\langle{e^{\frac{a W a}{4}}S_E}\rangle S_E\rho[\Gamma]\, ,
\ee
and such that the mean value \(\langle e^{\frac{a W a}{4}} S_E\rangle\) is different from \(0\). The correlation matrix
\be
\Gamma=1-\frac{\langle e^{\frac{a W a}{4}}S_E a\otimes a\rangle}{\langle e^{\frac{a W a}{4}} S_E\rangle}
\ee
can be used to perform any product at the price to have the string \(S_E\).
We observe that
\be
S_E\rho[\Gamma]=\rho[P_E\Gamma P_E]S_E\, ,
\ee
where \(P_E\) is the diagonal operator with elements \(i=-1\) if \(i\in E\) and \(1\) otherwise. 
In fact, exploiting the commutation relation, products of operators like \(Q_W\) can be carried to the form 
\be
\prod_{i=1}^m Q_{W_i}=\prod_{i=1}^m\langle{e^{\frac{a W_i a}{4}}S_{E_i}\rangle}\prod_{i=1}^m\rho[\Gamma_i]S_{\tilde E}\, ,
\ee
where \(S_{\tilde E}\) is the final \(\sigma_z\) string.
The trace 
\bea
\fl
\Tr{\prod_{i=1}^m Q_{W_i}}&=&\prod_{i=1}^m\langle{e^{\frac{a W_i a}{4}}S_{E_i}}\rangle\Tr{\prod_{i=1}^m\rho[\Gamma_i]S_{\tilde E}}= \nonumber\\ \fl &=&
\prod_{i=1}^m\langle{e^{\frac{a W_i a}{4}}S_{E_i}}\rangle\{\Gamma_1,\cdots,\Gamma_m\}\Tr{\rho[\Gamma_1\times\cdots\times\Gamma_m]S_{\tilde E}}
\eea
is free from pathologies.


\section*{References}

\begin{thebibliography}{99}

\bibitem{Renyi} L Amico, R Fazio, A Osterloh, and V Vedral, Entanglement in many-body systems,
Rev. Mod. Phys. {\bf 80}, 517 (2008);
J Eisert, M Cramer, and M B Plenio, Area laws for the entanglement entropy - a review,
Rev. Mod. Phys. {\bf 82}, 277 (2010);
Entanglement entropy in extended systems,
P Calabrese, J Cardy, and B Doyon Eds, J. Phys. A {\bf 42} 500301 (2009).
%
\bibitem{cl-08}
P Calabrese and A Lefevre,
Entanglement spectrum in one-dimensional systems,
Phys. Rev. A {\bf 78}, 032329 (2008).
%
\bibitem{mps}
N Schuch M M Wolf, F Verstraete, and J I Cirac, Entropy scaling and simulability by matrix product states,
Phys. Rev. Lett. {\bf 100}, 030504 (2008);
D Perez-Garcia, F Verstraete, M M  Wolf, J I Cirac, Matrix Product State Representations
Quantum Inf. Comput. {\bf 7}, 401 (2007);
L Tagliacozzo, T R. de Oliveira, S Iblisdir, and J I Latorre,
Scaling of entanglement support for Matrix Product States,
Phys. Rev. B {\bf 78}, 024410 (2008);
F Pollmann, S Mukerjee, A M Turner, and J E Moore,
Theory of finite-entanglement scaling at one-dimensional quantum critical points,
Phys. Rev. Lett. {\bf 102}, 255701 (2009);
F Verstraete and J I Cirac, Renormalization and tensor product states in spin chains and lattices,
J. Phys. A {\bf 42} 504004 (2009).

%
\bibitem{Holzhey}  C Holzhey, F Larsen, and F Wilczek,
Geometric and renormalized entropy in conformal field theory, Nucl. Phys. B {\bf 424}, 443 (1994).
%
\bibitem{cc-04}
P Calabrese and J Cardy,
Entanglement entropy and quantum field theory, J. Stat. Mech. P06002 (2004).
%
\bibitem{cc-rev}
P Calabrese and J Cardy,
Entanglement entropy and conformal field theory,
J. Phys. A {\bf 42}, 504005 (2009).
%

\bibitem{ch-08}
H. Casini, C. D. Fosco, and M. Huerta,
Entanglement and alpha entropies for a massive Dirac field in two dimensions,
J. Stat. Mech. P05007 (2005);
H. Casini and M. Huerta,
Remarks on the entanglement entropy for disconnected regions,
JHEP 0903: 048 (2009);
H. Casini and M. Huerta,
Reduced density matrix and internal dynamics for multicomponent regions, 
Class. Quant. Grav. {\bf 26}, 185005 (2009).


\bibitem{ffip-08}
P. Facchi, G. Florio, C. Invernizzi, and S. Pascazio,
Entanglement of two blocks of spins in the critical Ising model,
Phys. Rev. A {\bf 78}, 052302 (2008).

\bibitem{kl-08} I Klich and L Levitov,
Quantum noise as an entanglement meter,
Phys. Rev. Lett. {\bf 102}, 100502 (2009).


\bibitem{cg-08}
M Caraglio and F Gliozzi, Entanglement entropy and twist fields,
JHEP 0811: 076 (2008).

\bibitem{fps-09}
S Furukawa, V Pasquier, and J Shiraishi,
Mutual information and compactification radius in a c=1 critical phase in
one dimension, Phys. Rev. Lett. {\bf 102}, 170602 (2009).

\bibitem{cct-09}
P Calabrese, J Cardy, and E Tonni,
Entanglement entropy of two disjoint intervals in conformal field theory,
J. Stat. Mech. P11001 (2009).

\bibitem{atc-09}
V Alba, L Tagliacozzo, and P Calabrese,
Entanglement entropy of two disjoint blocks in critical Ising models, 
Phys. Rev. B {\bf 81} (2010) 060411.

\bibitem{Dixon}
L J Dixon, D Friedan, E J Martinec and S HShenker,
The conformal field theory of orbifolds,
Nucl. Phys.  B {\bf 282} (1987) 13;
Al. B. Zamolodchicov, Conformal scalar field on the hyperelliptic curve 
and critical Ashkin-Teller multipoint correlation functions,
Nucl. Phys. B {\bf 285} (1987) 481;
V. G. Knizhnik, Analytic fields on Riemann surfaces. II,
Communn. Math. Phys. {\bf 112}, 567 (1987);
M. Bershadsky and A. Radul,
Conformal field theories with additional $Z_N$ symmetry,
Int. J. Mod. Phys. A {\bf 2}, 165 (1987).


\bibitem{Neg}
H. Wichterich, J. Molina-Vilaplana, and S. Bose,
Scale invariant entanglement at quantum phase transitions,
Phys. Rev. A {\bf 80}, 010304(R) (2009).

\bibitem{Neg2}
S. Marcovitch, A. Retzker, M. B. Plenio, and B. Reznik,
Critical and noncritical long range entanglement in the Klein-Gordon field,
Phys. Rev. A {\bf 80}, 012325 (2009).

\bibitem{Neg3}
H. Wichterich, J. Vidal, and S. Bose,
Universality of the negativity in the Lipkin-Meshkov-Glick model, 0910.1011.

\bibitem{ip-09}
F Igloi and I Peschel, On reduced density matrices for disjoint subsystems, 0910.5671
%


\bibitem{afc-09}
V Alba, M Fagotti, and P Calabrese, Entanglement entropy of excited states,
J. Stat. Mech. (2009) P10020.

\bibitem{xxz}
N. Kitanine, J. M. Maillet, and V. Terras, 
Form factors of the XXZ Heisenberg spin-1/2 finite chain 
Nucl. Physics B {\bf 554} 647 (1999);
N. Kitanine, J. M. Maillet, and V. Terras, 
Correlation functions of the XXZ Heisenberg spin-1/2 chain in a magnetic field,
Nucl. Phys. B {\bf 567}, 554 (2000);
J. Sato and M. Shiroishi, 
Density matrix elements and entanglement entropy for the spin-1/2 XXZ chain 
at $\Delta$=1/2,
J. Phys. A {\bf 40}, 8739 (2007);
J. Damerau, F. G\"ohmann, N. P. Hasenclever, and A. Kl\"umper,
Density matrices for finite segments of Heisenberg chains of arbitrary length 
J. Phys. A {\bf 40}, 4439 (2007);
H. E. Boos, J. Damerau, F. G\"ohmann, A. Kl\"umper, J. Suzuki, and A. Weisse,
Short-distance thermal correlations in the XXZ chain,
J. Stat. Mech. P08010 (2008);
C. Trippe, F. G\"ohmann, A. Kl\"umper,
Short-distance thermal correlations in the massive XXZ chain, Eur. Phys. J. B {\bf 73}, 253 (2009);
J. Sato, M. Shiroishi, M. Takahashi, 
Exact evaluation of density matrix elements for the Heisenberg chain ,
J. Stat. Mech. P12017 (2006);
H Katsura and I Maruyama,
Derivation of Matrix Product Ansatz for the Heisenberg Chain from Algebraic Bethe Ansatz,
0911.4215.



\bibitem{Vidal} G Vidal, J I Latorre, E Rico, and A Kitaev,
Entanglement in quantum critical phenomena,
Phys. Rev. Lett. {\bf 90}, 227902 (2003);
J I Latorre, E Rico, and G Vidal,
Ground state entanglement in quantum spin chains,
Quant. Inf. Comp. {\bf 4}, 048 (2004).


\bibitem{gl-rev} J I Latorre and A Riera,
A short review on entanglement in quantum spin systems,  J. Phys. A {\bf 42}, 504002 (2009);
I Peschel and V Eisler, Reduced density matrices and entanglement entropy in free lattice models,
J. Phys. A {\bf 42}, 504003 (2009).


\bibitem{jk-04}
B-Q Jin and V E Korepin,
Quantum spin chain, Toeplitz determinants and Fisher-Hartwig conjecture,
J. Stat. Phys. {\bf 116}, 79 (2004);
%
A R Its, B-Q Jin, and V E Korepin,
Entanglement in XY spin chain,
J. Phys. A {\bf 38}, 2975 (2005);
F Franchini, A R Its, and V E Korepin
Renyi entropy of the XY spin chain, J. Phys. A {\bf 41} (2008) 025302.

\bibitem{ccen-10} P Calabrese, M Campostrini, F Essler, and B Nienhuis,
Parity effects in the scaling of block entanglement in gapless spin chains, Phys. Rev. Lett. to appear 0911.4660.

\bibitem{if-09}
Kh D Ikramov and H Fasbender, On the product of two swek-Hamiltonians or two skew-symmetric matrices,   
J. Math. Sci. {\bf 157}, 697 (2009).

%

%
\bibitem{lsca-06}
N Laflorencie, E S Sorensen, M-S Chang, and I Affleck,
Boundary effects in the critical scaling of entanglement entropy in 1D systems,
Phys. Rev. Lett. {\bf 96}, 100603 (2006);
E S Sorensen, N Laflorencie, and I Affleck,
Entanglement entropy in quantum impurity systems and systems with boundaries,
J. Phys. A {\bf 42}, 504009 (2009).





\bibitem{r-2}
H-Q Zhou, T Barthel, J O Fjaerestad, and U Schollwoeck,
Entanglement and boundary critical phenomena,
Phys. Rev. A {\bf 74}, 050305 (2006);
G De Chiara, S Montangero, P Calabrese, and R Fazio,
Entanglement entropy dynamics in Heisenberg chains,
J. Stat. Mech. (2006) P03001;
O Legeza, J Solyom, L Tincani, and R M Noack,
Entropic analysis of quantum phase transitions from uniform to spatially inhomogeneous phases,
Phys. Rev. Lett. {\bf 99}, 087203 (2007);
A Laeuchli and C Kollath,
Spreading of correlations and entanglement after a quench in the
Bose-Hubbard model, J. Stat. Mech. (2008) P05018;
G Roux, S Capponi, P Lecheminant, and P Azaria,
Spin 3/2 fermions with attractive interactions in a one-dimensional optical lattice: phase diagrams, entanglement entropy, and the effect of the trap,
Eur. Phys. J. B {\bf 68}, 293 (2009);
B Nienhuis, M Campostrini, and P Calabrese,
Entanglement, combinatorics and finite-size effects in spin-chains,
J. Stat. Mech. (2009) P02063;
F Gliozzi and L Tagliacozzo, Entanglement entropy and the complex plane of replicas,
J. Stat. Mech. (2010) P01002;
I J Cirac and G Sierra, Infinite matrix product states, conformal field theory
and the Haldane-Shastry model, 0911.3029;
M B Hastings, I Gonzalez, A B Kallin, R G Melko,
Measuring Renyi Entanglement Entropy with Quantum Monte Carlo,
1001.2335;
J C Xavier, Entanglement entropy, conformal invariance and the critical behavior of the anisotropic
spin-S Heisenberg chains: A DMRG study, 1002.0531;
H F Song, S Rachel, and K Le Hur,
General relation between entanglement and fluctuations in one dimension, 1002.0825.

\bibitem{ar-10}
F C Alcaraz and V Rittenberg, Shared Information in Stationary States at Criticality, 0912.2963;
A B Kallin, I Gonz\'alez, M B Hastings, and R Melko,
Valence bond and von Neumann entanglement entropy in Heisenberg ladders,
Phys. Rev. Lett. {\bf 103} (2009) 117203.

\bibitem{cc-10}
J Cardy and P Calabrese, Unusual Corrections to Scaling in Entanglement Entropy, 1002.4353.

\bibitem{p-04}
I. Peschel, On the entanglement entropy for a XY spin chain, J. Stat. Mech. (2004) P12005.

\bibitem{ij-08}
F Igloi and R Juhasz,
Exact relationship between the entanglement entropies of XY and quantum Ising chains,
Europhys. Lett. {\bf 81}, 57003 (2008).


\bibitem{ccd-07}
J  L Cardy,  O A Castro-Alvaredo, and B Doyon,
Form factors of branch-point twist fields in quantum integrable models and
entanglement entropy,   J. Stat. Phys. {\bf 130} (2008) 129;
O. A. Castro-Alvaredo and  B. Doyon,
Bi-partite entanglement entropy in integrable models with backscattering,
J. Phys. A {\bf 41}, 275203 (2008);
O A Castro-Alvaredo and B Doyon,
Bi-partite entanglement entropy in massive 1+1-dimensional quantum field theories
J. Phys. A {\bf 42}, 504006 (2009).

\bibitem{cc-05} P. Calabrese and J. Cardy,
Evolution of Entanglement entropy in one dimensional systems,
J. Stat. Mech. P04010 (2005).

\bibitem{fc-08}
M Fagotti and P Calabrese, 
Evolution of entanglement entropy following a quantum quench: 
Analytic results for the XY chain in a transverse magnetic field,
Phys. Rev. A 78, 010306 (2008).

\bibitem{sc-08}
S Sotiriadis and J Cardy Inhomogeneous Quantum Quenches,  J. Stat. Mech. (2008) P11003.

\bibitem{ep}
V. Eisler and I. Peschel,
Evolution of entanglement after a local quench,
J. Stat. Mech. P06005 (2007);
V. Eisler, D. Karevski, T. Platini, and I. Peschel, 
Entanglement evolution after connecting finite to infinite quantum chains,
J. Stat. Mech. (2008) P01023.

\bibitem{rm-05}
G. Refael and J. E. Moore,
Entanglement entropy of random quantum critical points in one dimension,
Phys. Rev. Lett. {\bf 93}, 260602 (2004);
Scaling of Entanglement Entropy in the Random Singlet Phase,
Phys. Rev. B {\bf 72}, 140408(R) (2005);
R. Santachiara,
Increasing of entanglement entropy from pure to random quantum critical chains,
J. Stat. Mech. (2006) L06002;
G. Refael and J. E. Moore, 
Criticality and entanglement in random quantum systems,
J. Phys. A {\bf 42}, 504010 (2009).

\bibitem{p-89}
J Palmer, Pfaffian Bundles and the Ising Model, Commun. Math. Phys. {\bf 120}, 547 (1989).


%

\end{thebibliography}
\end{document}